\documentclass{emulateapj}

\def\teff{T$_{\rm{eff}}\,$}
\def\teffs{T$_{\rm{eff}}s\,$}

\def\etal{{~et~al.}}

\def\mj{$M_{\rm J}\,$}

\def\Dwa{$\,$\uppercase\expandafter{\romannumeral5}$\,$}
\def\mic{$\mu$m$\,$}

\def\sless{\lower2pt\hbox{$\buildrel {\scriptstyle <}
   \over {\scriptstyle\sim}$}}
\def\sgreat{\lower2pt\hbox{$\buildrel {\scriptstyle >}
   \over {\scriptstyle\sim}$}}
\def\aa{Astron. Astrophys.\ }

\slugcomment{Accepted to the Ap. J.}

\null\voffset=+0.0pc  

\begin{document}

\title{L and T Dwarf Models and the L to T Transition}

\author{Adam Burrows\altaffilmark{1}, David Sudarsky\altaffilmark{1}, \&  Ivan Hubeny\altaffilmark{1}}

\altaffiltext{1}{Department of Astronomy and Steward Observatory, 
                 The University of Arizona, Tucson, AZ \ 85721}

\begin{abstract}

Using a model for refractory clouds, a novel algorithm for handling 
them, and the latest gas-phase molecular opacities, we have produced a new 
series of L and T dwarf spectral and atmosphere models
as a function of gravity and metallicity, spanning the \teff\ range from 2200 K to 700 K.
The correspondence with observed spectra and infrared colors for early- and mid-L dwarfs
and for mid- to late-T dwarfs is good.  We find that the width in infrared color-magnitude
diagrams of both the T and L dwarf branches
is naturally explained by reasonable variations in gravity and, therefore, that gravity may be 
the ``second parameter" of the L/T dwarf sequence.  We investigate the dependence
of theoretical dwarf spectra and color-magnitude diagrams upon various cloud properties, such
as particle size and cloud spatial distribution.  In the
region of the L$\rightarrow$T transition, we find that
no one cloud-particle-size and gravity combination can be made to fit
all the observed data, though the role of binarity in the extant L and T dwarf
data sets has yet to be fully determined.  However, our results suggest  
that current ignorance of detailed cloud meteorology renders ambiguous the extraction
of various physical quantities such as \teff and gravity for mid-L to early-T dwarfs.
Nevertheless, for decreasing \teff, we capture with some accuracy the transition from CO
to CH$_4$ in the $K$ band, the FeH and CrH features at $\sim$1.0 \mic
and $\sim$0.85 \mic (at least for the early- to mid-Ls), the 
emergence of the neutral Na, K, Cs, and Rb alkali features,
the growth of the $Y$/$Z$ band peak, the disappearance of TiO and VO in L dwarfs,
the evolution in the shapes of the $J$, $H$, and $K$ bands, and the
evolution of the K I features near 1.25 \mic\ and 1.17 \mic. We
speculate that the subdwarf branch of the L dwarfs will be narrower 
in effective temperature, and, that for low enough
metallicity, the L dwarfs will disappear altogether as a spectroscopic class.  Furthermore,
we note that the new, lower solar oxygen abundances of Allende-Prieto, Lambert, \& Asplund (2002)
produce better fits to brown dwarf data than do the older values.   
Finally, we discuss various issues in cloud physics and modeling and speculate on how a 
better correspondence between theory and observation in the 
problematic L$\rightarrow$T transition region might be achieved.

\end{abstract}

\section{Introduction}
\label{intro}

In the last ten years since the first identification of T and L dwarfs in the solar neighborhood,
more than sixty T dwarfs and hundreds of L dwarfs have been classified spectroscopically and photometrically
(Oppenheimer et al. 1995; Delfosse et al. 1997; Burgasser et al. 1999,2000a,b,c; 
Kirkpatrick et al. 1999,2000; Geballe et al. 2002; Mart{\'{\i}}n et al. 
1999; Cruz et al. 2003; Knapp et al. 2004; Golimowski et al. 2004).  
Both infrared and optical spectral features and colors have been used to define subtypes 
from L1 to L9 and from T0 to T9.  These subtypes span an effective temperature (\teff) range
from $\sim$2300 K to $\sim$750 K and encompass the edge of the hydrogen-burning main sequence
(\teff $\sim$ 1700 K; L5$-$L6).  Water (H$_2$O) bands dominate the spectra 
of both Ls and Ts.  However, the L dwarfs are characterized by the disappearance  
of the optical bands of TiO and VO (Jones \& Tsuji 1997) that define 
the M dwarfs, the growth of near-infrared 
FeH and CrH features, the emergence of CH$_4$ in the $K$ ($\sim$2.0$-$2.2 \mic) 
and L$^{\prime}$ ($\sim$3.3 \mic) bands, and
the progressive reddening of their $J-K$ colors.  The T dwarfs are distinguished 
by the appearance of strong CH$_4$ features in the $H$ ($\sim$1.5$-$1.8 \mic) 
and the $K$ bands, the weakening and disappearance of
the metal hydride features, the growing importance of collision-induced absorption by
H$_2$ (Borysow \& Frommhold 1990; Borysow, J\o{rgensen}, \& Zheng 1997), 
the strength of neutral alkali metal resonance lines shortward of 1.0 micron 
(Burrows, Marley, \& Sharp 2000), and  ``anomalously" blue infrared colors.  It is thought 
that the L/T transition is near $\sim$1300 K. All T dwarfs are brown dwarfs, with substellar masses, 
but many L dwarfs, however exotic, could be stars with masses above the
main sequence edge. 

While most of these trends and characteristics are understood in terms of general chemistry
(Burrows \& Sharp 1999; Fegley \& Lodders 1996; Lodders 1999), the centrality of molecules 
in L and T dwarf atmospheres has put a new premium on the calculation of molecular opacities and abundances.
Though theorists have gradually responded to this challenge 
(Partridge \& Schwenke 1997; Burrows et al. 2002a; Dulick et al. 2003; Burrows \& Volobuyev 2003),
their remain significant deficits in the relevant molecular databases.  A glaring example 
is the absence of data on the hot bands of methane in the $H$ band, but across the board
molecular spectroscopy lags the atomic spectroscopy developed over the last hundred years to
support classical stellar atmosphere calculations.  As a result, theoretical models of
L and T dwarf atmospheres are still evolving and the extraction of physical quantities such as
\teff, gravity ($g$), and metallicity from the growing library of well-calibrated spectra, though
possible in a crude sense, remains imprecise (Burrows et al. 2002b; Marley 
et al. 2002; Tsuji, Nakajima, \& Yanagisawa 2004; Burgasser, Burrows, \& Kirkpatrick 2005).

However, the most difficult challenge confronting theorists designing models of L dwarfs
and the L to T transition is the accurate incorporation of condensate clouds ({\it e.g.,} 
calcium aluminates, silicates, iron). Such refractories form at the M to L transition
temperatures ($\sim$2300$-$1800 K), persist and thicken to dominate and define 
the L dwarfs, and when their atmospheric effects 
finally diminish are thereby responsible for the L to T transition 
(Burrows et al. 2001; Ackermann \& Marley 2001; Marley et al.
2002; Tsuji, Ohnaka, \& Aoki 1999).  Were it not for condensate clouds
the Ls would not be red in infrared colors and get redder with 
decreasing \teff. In fact, the L dwarfs would not exist as a spectroscopic type 
\footnote{This fact serves to emphasize that the span and importance of the L dwarf family is a function
of metallicity, and for subdwarfs will be narrower in \teff.}.
Hence, to understand the M$\rightarrow$L$\rightarrow$T 
sequence and their spectra in detail requires a mastery of not only the chemical
condensation sequences and the consequent elemental depletions (``rainout"; Burrows \& Sharp 1999)
in a gravitational field, but the spatial extent, particle size and shape distributions,
grain optical properties, and meteorology of clouds as well.  These complications do not confront 
one who models most other types of ``stars."

After the first rush of basic discoveries, observers are now beginning
to refine our knowledge of L and T dwarfs from the optical to the mid-infrared.
Some target facts to explain include the brightening in the $J$ ($\sim$1.2$-$1.3 \mic) 
and $Y$/$Z$ ($\sim$1.0$-$1.1 \mic) bands
for the early T dwarfs (Dahn et al. 2002; Tinney, Burgasser, \& 
Kirkpatrick 2003; Vrba et al. 2004), despite the 
decrease in bolometric luminosity along the spectroscopic sequence (Golimowski et al. 2004),
the resurgence in the early Ts of FeH at $\sim$0.99 \mic (McLean et al. 2003; 
Burgasser et al. 2004; Nakajima, Tsuji, \& Yanagisawa 2004; Cushing, Rayner, \& Vacca 2005),
the non-equilibrium CO abundances inferred from spectral observations at 4.5$-$4.9 \mic
(Golimowski et al. 2004), the appearance of the 2.2 \mic and 3.3 \mic CH$_4$ features in
the late Ls, and the narrow apparent range of \teff at the M$\rightarrow$L and L$\rightarrow$T
transitions. In fact. Golimowski et al. (2004) suggest that \teff is nearly constant from L7 to T4.5.
Moreover, the actual prevalence of spectroscopic binaries among
the observed L and T dwarfs, and the resulting misplacements on the HR diagram,
remain a concern and a caution (Liu \& Leggett 2005; Gelino \& Kulkarni 2005; Burgasser et al. 2005). 

The brightening in the $J$ band is the most intriguing anomaly. Its explanation must 
be the rapid thinning out of the clouds in the spectrum-forming region of the brown dwarf atmosphere
during the L$\rightarrow$T transition.  But how the effective opacity
of the silicate clouds decreases so quickly with \teff and spectroscopic subtype 
to yield heavy-element depleted T dwarf atmospheres has yet to be explained. 
Ideas include the emergence of holes in the clouds, whose areal filling factor
would need to increase with spectral subtype (Burgasser et al. 2004), 
a sudden collapse of the cloud deck (Tsuji \& Nakajima 2003;
Tsuji, Nakajima, \& Yanagisawa 2004), or an increase in the ``sedimentation efficiency"
of the clouds, with a concommitent rapid increase in the silicate particle size
(Knapp et al. 2004).  All these models have their strengths and weaknesses,
and all are ad hoc, with little physical or meteorological justification for the rapid 
transitions proffered. 

There is emerging evidence that the L and T spectroscopic sequences are not
a one-parameter family in \teff, but that there is a second (and, perhaps, a third)
parameter.  Burrows et al. (2002b) and Knapp et al. (2004) suggest that this 
second parameter is gravity and our calculations here lend strong support to this notion.
The widths of both the L and the T dwarf regions in HR diagrams are best explained
with a range of gravities.  Metallicity may also play a role, but a subordinate one,
except for rare outliers, and we explore in this paper the metallicity dependence
of L and T dwarf model atmospheres, spectra, and colors.

In this paper, we provide a new generation of L dwarf and T dwarf model spectra, colors,
and atmospheres.  We use updated FeH, CrH, TiO, and VO gas-phase opacities and abundances and investigate
the dependence on cloud model parameters, in particular particle size and cloud extent/shape. 
To do so, we have developed a sophisticated and self-consistent algorithm for handling clouds in
atmosphere calculations.  With this algorithm, we demonstrate the ambiguities that
remain in any treatment of the L$\rightarrow$T transition, while generating 
good spectral and color models for both the early- and mid-L dwarfs 
and the mid- and late-T dwarfs.  In the process, we provide and discuss 
new spectral models, temperature/pressure profiles, color-magnitude diagrams, and color-color
diagrams for a broad range of \teffs, gravities, metallicities, and particle sizes.

In \S\ref{database}, we list the numerical
tools and databases we have used to generate self-consistent molecular atmospheres
and discuss the numerical method we employ to handle clouds and convection in 
L and T dwarf atmospheres.  We review in \S\ref{model} general cloud physics and our 
cloud model parametrizations. Then, we continue in \S\ref{param} with a discussion of  
the effects of cloud shape, particle size, and gravity. 
The associated models are provided only to explore parameter dependences;   
they are not our best model results. We then introduce our baseline model,
which, though manifestly limited given the current rudimentary knowledge of  
cloud physics and incorporating as it does simple ansatze (as do all other published 
models), nevertheless gives acceptable fits to the observations
outside of the problematic L9-T3 spectral range.  This model suite
is being provided to astronomers to aide them in understanding 
the rich harvest of L and T dwarf spectral data.   
In \S\ref{profiles} we discuss representative temperature/pressure ($T/P$) 
profiles that serve as the jumping-off point for further
speculation on the L$\rightarrow$T transition.  Next, in \S\ref{spectra} we provide spectra for our fiducial model set
as a function of \teff, gravity, and metallicity, as well as a few comparisons between our spectral models and data to
establish that the models fit reasonably well.   This is followed in \S\ref{col-mag} by
M$_J$ versus $J-K$, M$_K$ versus $J-K$, and M$_{z^{\prime}}$ versus $i^{\prime} - z^{\prime}$ color-magnitude 
diagrams for these baseline models.  For comparison, data for M, L, and T dwarfs with 
known parallaxes are superposed.  Finally, in \S\ref{conclusions} we summarize 
our conclusions and discuss what remains to be done.

\section{Numerical Tools and Chemical Databases}
\label{database}

In a study such as the one we conduct for this paper many self-consistent spectra and 
atmospheres must be generated for a variety of physical parameters.  To do so expeditiously,
it is undesirable and expensive to calculate ``on the fly" the opacities at a given wavelength
from the extensive line lists available. Rather, we precompute and tabulate at 5000 frequency points,
logarithmically spaced from 0.3 \mic to 300 \mic, the absorption and scattering 
opacities on a mesh in $T$/$P$ space for $\sim$30 
of the dominant molecular and atomic species found in L and T dwarf atmospheres.  These
tables are then used with a table of the corresponding molecular and atomic abundances 
at the desired metallicity to produce a large table of total opacities as a function of 
frequency, temperature, and pressure.  Such a table fully incorporates the physics 
of line broadening and stimulated emission and is interpolated in during the iteration
of the spectral/atmosphere model until a converged $T/P$ profile and a spectrum in radiative equilibrium are obtained.
Instead of requiring hours on a supercomputer to generate a model, this procedure typically requires only
minutes on a single-processor workstation.  To obtain spectra at higher frequency resolution, the
derived $T/P$ profile can be reused with a higher-resolution opacity table, without the need to iterate
further.

The atmospheric structure is computed using a specific variant of the 
stellar atmospheres code TLUSTY (Hubeny 1988; Hubeny 1992; Hubeny \& Lanz 1995),
with modifications as described in Burrows et al. 2002b, Sudarsky, Burrows, \& Hubeny (2003)
and Hubeny, Burrows, \& Sudarsky (2003). We have developed a computational scheme to treat
clouds and their interplay with convection zones efficiently. The approach
will be described in detail elsewhere (Hubeny et al., in prep.);
here we briefly mention some of its most important aspects.

We have reformulated the general linearization scheme
in the so-called Rybicki-type approach (Mihalas 1978),
which consists in writing the global linearization (Jacobi)
matrix in block form, where each block contains an
individual state parameter for all depth points.
This allows us to avoid using
Accelerated Lambda Iteration (ALI) to treat the radiative
transfer equation.  Instead, we use full linearization of the
radiation intensities, which leads to much faster and more stable
convergence, without compromising on computer time.

Convection is treated in the mixing-length formalism,
with a mixing length of one pressure scale height. In
actual brown dwarf atmospheres, the temperature gradient
is very close to the adiabatic gradient. During the linearization iteration,
it often happens that the current gradient is somewhat lower
than the adiabatic gradient, even when in reality it should be larger.
We have devised a procedure that detects such spurious ``convection gaps" and recomputes the
temperature structure in such a way that at those points the convective flux is
indeed non-zero before entering the next iteration
of the linearization.

In addition to introducing a flexible way to treat cloud shapes (see \S\ref{struct}),
we have developed several ways to self-consistently determine the cloud's position.
It turns out that it is more stable not to recalculate the cloud position after each iteration; instead it
is preferrable to perform several (2$-$4) iterations, keeping the positions
of the clouds fixed between updates. In addition, we 
have introduced a flexible depth-point rezoning scheme.
Essentially, we place additional depth points close to the
cloud base whenever the new cloud position is recalculated.
Finally, in some cases the cloud position is found to oscillate
between two (sometimes even three) distinct positions. This problem
is avoided by setting the updated position of the cloud base between 
the previous position of the cloud base and the intersection
of the T/P profile and the condensation curve.
At convergence, the cloud base will be at the
proper intersection point.

The equation of state from which we obtain the $T/P$/density($\rho$) relation is that of
Saumon, Chabrier, \& Van Horn (1995).  The molecular compositions 
and ionization fractions (Burrows, Sudarsky, \& Lunine 2003) are obtained using
the chemical equilibrium code SOLGASMIX (Burrows and Sharp 1999), with updated
thermochemical data for Ti and V compounds, silicates and calcium aluminates, H$^{-}$, and metal hydrides. 
The abundance code incorporates prescriptions for rainout and depletion due to
condensate formation of refractory silicates, aluminates, titanates, iron, water, and ammonia.
Despite the overall complexity of the problem, the most important atoms and non-refractory molecules 
comprise a small subset: H$_2$, H$_2$O, CH$_4$, CO, N$_2$, and NH$_3$, FeH, CrH, TiO, VO,
Na, and K.

Our always-evolving opacity/spectroscopy database would require a paper in itself to explain,
but is partially described (with references) in Burrows et al. (1997,2001,2002a,2003).
The opacities of the alkali metal atoms are taken from Burrows, Marley, \& Sharp (2000),
which are similar in the line cores and near wings to those found in Burrows \& Volobuyev (2003).
For future work, we have extended the opacity database into the UV down to $\sim$0.08 \mic.
Our solar-metallicity elemental abundances are taken from
Allende-Prieto, Lambert, \& Asplund (2002), which supercedes Anders \& Grevesse (1989).
Allende-Prieto, Lambert, \& Asplund have found that in the past the solar abundances of oxygen and carbon were
overestimated by as much as $\sim$35$-$45\%.  While this change has but subtle
consequences for stellar and solar atmospheres, since the water abundance so dominant in L and T dwarf atmospheres
scales almost directly with the oxygen abundance, these atmospheres are much more directly affected by the change.
Using Spitzer IRAC data and our theoretical models, Patten et al. (2004) show that the
new abundances fit the mid-infrared data much better.  Hence, and in a curious and unexpected way,
the new solar abundances are partially validated by brown dwarf observations.

Cloud physics, structure, character, and opacities play a major 
role in the M$\rightarrow$L transition, L dwarf atmospheres,
and the L$\rightarrow$T transition and so are discussed 
separately in the following section.

\section{Clouds and Cloud Model Description}
\label{model}

\begin{figure}
\plotone{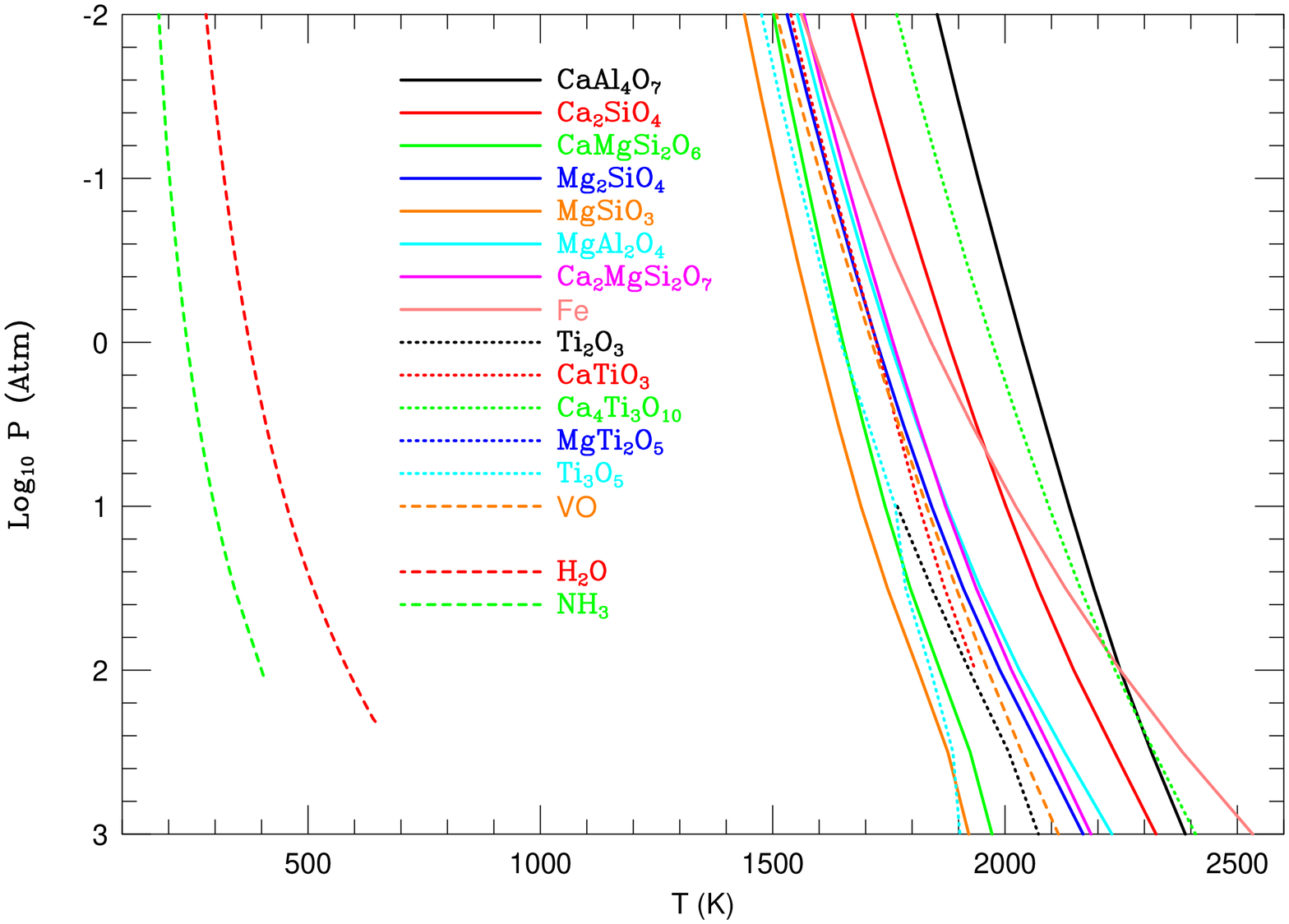}
\caption{Solar-metallicity condensation curves (temperature 
in Kelvin versus pressure in atmospheres) for many of the most important refractory species
thought to appear in the atmospheres of brown dwarfs and L dwarfs.  The most refractory
compound is the calcium aluminate grossite  
(CaAl$_4$O$_7$ $\equiv$ CaO + 2(Al$_2$O$_3$)).  Corundum (Al$_2$O$_3$), 
as such, does not generally form.  CaMgSi$_2$O$_6$ is diopside,  
Mg$_2$SiO$_4$ is forsterite, MgSiO$_3$ is enstatite, 
MgAl$_2$O$_4$ is spinel, and Ca$_2$MgSi$_2$O$_7$ is akermanite.
The dotted curves correspond to the refractory titanium compounds.  
Contrary to popular belief, condensed titanium is rarely in the form of perovskite (CaTiO$_3$).
The golden dashed curve is for condensed VO.  Liquid Fe is the solid peach curve at a slightly
shallower slope than those for the calcium/aluminum/magnesium condensates (solid colors).
Included for comparison are the condensation curves for water (H$_2$O, red dashed) and ammonia 
(NH$_3$, green dashed).  Notice how the condensation curves  
of the refractory compounds densely inhabit a narrow range of T/P space and that there is
a noticably wide gap between this refractory band and water. 
See text for a discussion of the salient features of this figure.}
\label{fig:1}
\end{figure}

In the atmospheres of late M dwarfs with \teff$\sim$2500 K, the most refractory 
compounds start to condense.  These are mostly the calcium aluminum oxides grossite
(CaAl$_4$O$_7$ $\equiv$ CaO + 2(Al$_2$O$_3$)) and hibonite (CaAl$_{12}$O$_{19}$
$\equiv$ CaO + 6(Al$_2$O$_3$), among others, and first appear in the upper
radiative, not the lower convective, zones.  Figure \ref{fig:1} depicts the
condensation curves of some of the most important refractory species thought to
form in late M dwarfs, L dwarfs, and brown dwarf atmospheres.  As Fig. \ref{fig:1}
indicates, this first phase of condensate cloud formation is followed, as the temperature drops, by the condensation
of titanium compounds, at the expense of gas-phase TiO, and later by the condensation of
VO.  Therefore, the disappearance of TiO and VO lags slightly, but follows close on the heals of,
the first emergence of refractory clouds.  Due to the low elemental abundances by number of titanium 
($\sim$8$\times$$10^{-8}$) and vanadium ($\sim$$10^{-8}$), the compounds 
into which TiO and VO condense (see Fig. \ref{fig:1}) are themselves not major components 
of this radiative cloud layer.   

In retrospect, this early transition from late M dwarfs to L dwarfs was first identified 
by Jones \& Tsuji (1997), who for late Ms noted the near 
simultaneous weakening of the TiO and VO spectral features
with the shallowing of the water troughs.  The latter is naturally explained by the appearance of a new
continuum absorber, clouds of the refractory calcium aluminates.  
However, though oxygen is abundant, the elemental abundances
of calcium and aluminum are not, just $\sim$2.0$\times$10$^{-6}$ and $\sim$2.8$\times$10$^{-6}$, respectively.
As a result, this first generation of clouds is rather thin, though the particle sizes in radiative
zones can be small ($\sim$0.5$-$5 microns, Cooper et al. 2003).  It is 
not until the appearance at slightly lower temperatures of condensates
containing silicon, magnesium, and/or iron, with elemental abundances 
\footnote{$\sim$3.3$\times$10$^{-5}$, $\sim$3.5$\times$10$^{-5}$, and 
$\sim$3.0$\times$10$^{-5}$, respectively} $\sim$10 times those
of calcium and aluminum, that the areal mass density of refractory clouds can be respectable.
With the lowering of \teff\ and the appearance of condensed iron and silicates 
(such as forsterite, Mg$_2$SiO$_4$, and enstatite, MgSiO$_3$), the photospheric layer
begins to coincide both with the thickening silicate cloud layer and with a convective zone,
and the pattern of the mid-L dwarf atmosphere is firmly established.  Therefore, the transitions
from M to L dwarf, from calcium aluminate to silicate/iron condensates, and from radiative
to convective clouds overlap, though not perfectly.

As Fig. \ref{fig:1} indicates, at a given pressure the many condensates can appear
in a narrow range of temperature.  Furthermore, whatever the particle sizes and optical
constants, the optical depths of such clouds are themselves sufficient after the early Ls to trip convection
(and the associated updrafts and downdrafts) where there are clouds.  Hence, 
after the early Ls (for which the first condensates inhabit a stably-stratified 
radiative zone) every condensate whose condensation curve
intersects the dwarf's $T/P$ profile will most probably reside in a common convection zone.  
It does not make sense to assume that each condensate is a separate, isolated layer, like a 
``pousse caf\'e." Rather, as the \teff\ decreases and the first clouds thicken, tripping
convection in the atmosphere, condensation and grain growth will 
occur over the same finite time and pressure/radius range
for many condensables.  The kinetics of such a soup of growing condensates poses
a daunting problem never before addressed, and not addressed here, though grain kinetics
in the brown dwarf context has been receiving some attention of late 
(Helling et al. 2001, 2004; Woitke \& Helling 2003,2004).  Nevertheless,
the equilibrium particle size distribution, achieved through the balance of
growth processes in the convective zone and grain destruction at and below 
the $T/P$-profile/condensation-curve intercept, is poorly constrained 
by theory (Cooper et al. 2003; Ackerman \& Marley 2001). Furthermore, the optical 
constants of heterogeneous grains of indeterminate composition
and layering are not easily derived from first principles.  Qualitatively, it is clear
that the modal particle size of grains in stable radiative zones is smaller ($\sim$0.1$-$5.0 \mic) 
than in turbulent convective zones ($\sim$10$-$150 \mic) (e.g., Cooper et al. 2003), 
but confidence in the current analytic estimates should not be great.

Once formed, clouds will not extend throughout the atmosphere above (to lower pressures and temperatures)
the condensation intercept, but will settle into a tighter spatial distribution bounded (approximately) from
below by this intercept.  Hence, the upper atmosphere will be depleted of the heavy elements
(such as Si, Mg, Fe, Ca, and Al) of which condensates are comprised.  The T dwarfs themselves
clearly indicate the general correctness of this conclusion; their clear spectra demand
that their condensate clouds reside at great depths below the formation region of the
emergent spectrum (Marley et al. 1996; Burrows et al 2002b).  The same is
true of Jupiter and Saturn, whose atmospheres are devoid of heavy
elements (Fegley \& Lodders 1996), but which must house at depth ($\sim$1000$-$3000 bars) both silicate 
and iron clouds.  The oxygen fraction in an L or T dwarf is so large
($\sim$5$\times$10$^{-4}$, solar) that condensation depletes its upper-atmospheric abundance
by no more than $\sim$15\% (Burrows \& Sharp 1999).  Below the cloud base (at higher pressures) 
is the ``infinite" reservoir of heavy elements that extends throughout the dwarf and which 
sets the heavy-element abundance boundary condition at the cloud base.  However, the heavy elements
that may have once existed in the upper atmosphere before condensation do not all remain
in the cloud once formed.  How much cloud material does remain in the cloud depends on the dynamics
of the cloud itself.  The actual areal mass density, spatial extent, and shape of a
cloud is not easily determined and depends upon the interplay of turbulent gas motions, grain growth,
overshoot, turbulent and eddy diffusion, and gravitational settling, as does the 
grain size distribution itself.  Ackerman \& Marley (2001) and Cooper et al. (2003)
suggest that the thickness of a cloud scales with the local pressure scale height
and that the cloud material is distributed exponentially from its base with a scale height between
$\sim$0.3 and $\sim$1.0 pressure scale heights. This seems reasonable, but the effect of multiple
condensation intercepts for the different species and of an extended convection zone that can encompass
these multiple condensation lines is not at all clear.  Nor is it clear how the modal particle
size varies with altitude, though there are some hints (Cooper et al. 2003).

For the time being, this state of complexity demands simple parametrizations.
Our approach has been to parametrize the spatial extent and shape of the condensate cloud
and to test various composite optical properties and modal particle sizes, loosely guided by
extant theory and previous practice (Cooper et al. 2003; Burrows et al. 1997,2001; Ackerman \& Marley 2001;
Allard et al. 2001) \footnote{For a given modal particle size, we use the Deirmendjian (1964)
particle size distribution.}.  For a given parametrization, we have compared the resulting
color-magnitude diagrams and spectra with observations in an attempt to settle on the fiducial
condensate-cloud approach that can be used for our next generation of atmosphere/spectral models
(published herein) for \teffs\ from 2200 K to 700 K and a variety of gravities and metallicities,  
collectively encompassing the L and T dwarf ranges (This is our model E described 
in \S\ref{struct}).  Since the mid- to late-Ts are not
affected by such clouds, this effort is germane only to L dwarf models 
and the L$\rightarrow$T transition.  In the process, we have determined the dependence
of dwarf spectra and colors on particle size and cloud morphology.

\subsection{Cloud Parametrizations}
\label{struct}

We assume that the clouds are uniformly distributed over the sphere 
and that there are no holes in either latitude or longitude.
The position of the high-pressure base of the cloud is set at the intersection of the
$T/P$ profile and the condensation curve. The corresponding
quantities are denoted by $P_0$ and $T_0$, and the corresponding column mass by $m_c$. 
(The pressure at a given depth is equal to $g\times{m}$, where $g$ is the surface gravity
and the areal mass column density is $m$.) The opacity (per gram of atmospheric material) of a condensate $a$ 
is given by
\begin{equation}
\kappa_a(\nu, m) = N_a\, M_a\, \frac{A_N}{\mu}\, S_a\, \sigma_a(\nu, a_0)\, f(m)\, ,
\end{equation}
where $N_a$ is the mixing ratio of species $a$, $M_a$ is its molecular weight 
(in grams), $\mu$ is the mean molecular weight of the atmospheric material, 
$A_N$ is Avogadro's number, $S_a$ is the supersaturation ratio,  
$\sigma_a(\nu, a_0)$ is the opacity per gram of species $a$ for
frequency $\nu$ and modal particle size/radius $a_0$, and 
$f(m)$ is our cloud shape function, defined such that the 
full presence of a cloud is denoted by $f=1\,$
while its complete absence is denoted by $f=0$. 

$N_a$ and $M_a$ are taken as known quantities (depending on
the assumed composition and varying with metallicity). The supersaturation
parameter and the modal particle size are taken as input parameters
of the model; for iron $S_{a}$ is 0.01 and for the silicates such as forsterite it is 1.0.
The cross sections for both true absorption and scattering are taken from
precalculated Mie tables, interpolating to the assumed particle size $a_0$.

In the absence of an ab-initio theory of cloud formation and merger, we envisage
a simple empirical model of a composite cloud having a flat part just
below the intersection point (upper cloud deck), and having an exponential decline
on the both sides of this flat part. The extent of the flat part, and the exponents
of the decline on both sides are free parameters of the problem. 
The cloud shape function is thus given by
\begin{eqnarray}
f(m) = (m/m_1)^{s_1}\, ,\quad\quad m \leq m_1\,  ,\nonumber\\
f(m) = 1 \, ,\quad\quad m_1 \leq m \leq m_0\, ,\nonumber\\
f(m) = (m/m_0)^{-s_2}\, , \quad\quad m \geq m_0\, ,
\label{flat}
\end{eqnarray}
where $m_1$ and $m_0$ are the upper (low-pressure) and lower (high-pressure) 
boundaries of the flat part of the cloud. $s_1$ and $s_2$ are indices 
such that $1/s_1$ and $1/s_2$ times the local pressure scale heights are the scale heights 
of the particle number distributions in the exponential tails.  $m_1$ can be parameterized in different
ways.  For a single isolated cloud species, $m_1$ would naturally be $m_c$
and the flat part would shrink to zero extent.   $1/s_1$ times the pressure 
scale height would indicate the cloud's approximate extent.  However, for multiple cloud condensates
with a range of $T_0$s that inhabit a convectively mixed region, the flat portion
should encompass the multiple cloud $T/P$ intercept points, with an extent roughly indicated
by the thickness of the condensate band in Fig. \ref{fig:1}.  This is why we have introduced
the flat portion in eq. (\ref{flat}).

To study the influence of the cloud shape function on the temperature structure
and the predicted spectra, we have considered the following five
cases. In model A, the cloud declines exponentially below
the high-temperature intersection point, but extends upward all the way to the surface ($s_1 = 0$; $s_2 = 10$).
Model B corresponds to having the cloud everywhere in the atmosphere ($s_1 = 0$; $s_2 = 0$).
Models A and B are similar and roughly correspond to the ``Dusty" models of 
Chabrier et al. (2000) and Allard et al. (2001).

Model C corresponds to a cloud decaying above and below the intersection
point without having a flat part ($s_1 = 6$; $s_2 = 10$); model D represent a cloud that decays
above the intersection point ($s_1 = 6$), while extending all the way down deep into the
dwarf ($s_2 = 0$); and, finally, model E corresponds to our generic, fiducial cloud -- a cloud with a flat part between
the intersection point of one of the most refractory condensates (such as grossite) and the intersection 
point for one of the less refractory condensates (such as forsterite), 
and exponential decays on both sides ($s_1 = 6$; $s_2 = 10$).
$s_1 = 6$ is steep, but comports with the expected thinness of single
clouds of refractory condensates having large particle radii.
Models E and D are similar.

Our default value of $m_0$ for the composite cloud model E is where $T_0$ for the  
most refractory condensates (the calcium aluminates) equals the atmospheric temperature. 
For simplicity, in our fiducial model E we have set $T_0$ equal 
to a constant 2300 K (Fig. \ref{fig:1}). Our $m_1$ for all models is taken to coincide with 
the intercept with the forsterite condensation curve (the forsterite $m_c$, at a temperature 
between 1500 and 1900 K; see Fig. \ref{fig:1}).  

Hence, the remaining parameters of our cloud models are the modal particle radius ($a_0$),
and the optical properties of the representative condensate.
We have studied the consequences of using enstatite, corundum, and forsterite for the latter
and have settled on forsterite (Mg$_2$SiO$_4$) for this generation of 
dwarf models.  Its abundance should be one of the largest in the mix,
and, as we will see (\S\ref{spectra}), when the particle radii are near $\sim$50$-$100 \mic,
the spectral fits over most of the relevant \teff and gravity 
range are rather good.

\section{Exploration of Parameter Dependences}   
\label{param}

\begin{figure}
\plotone{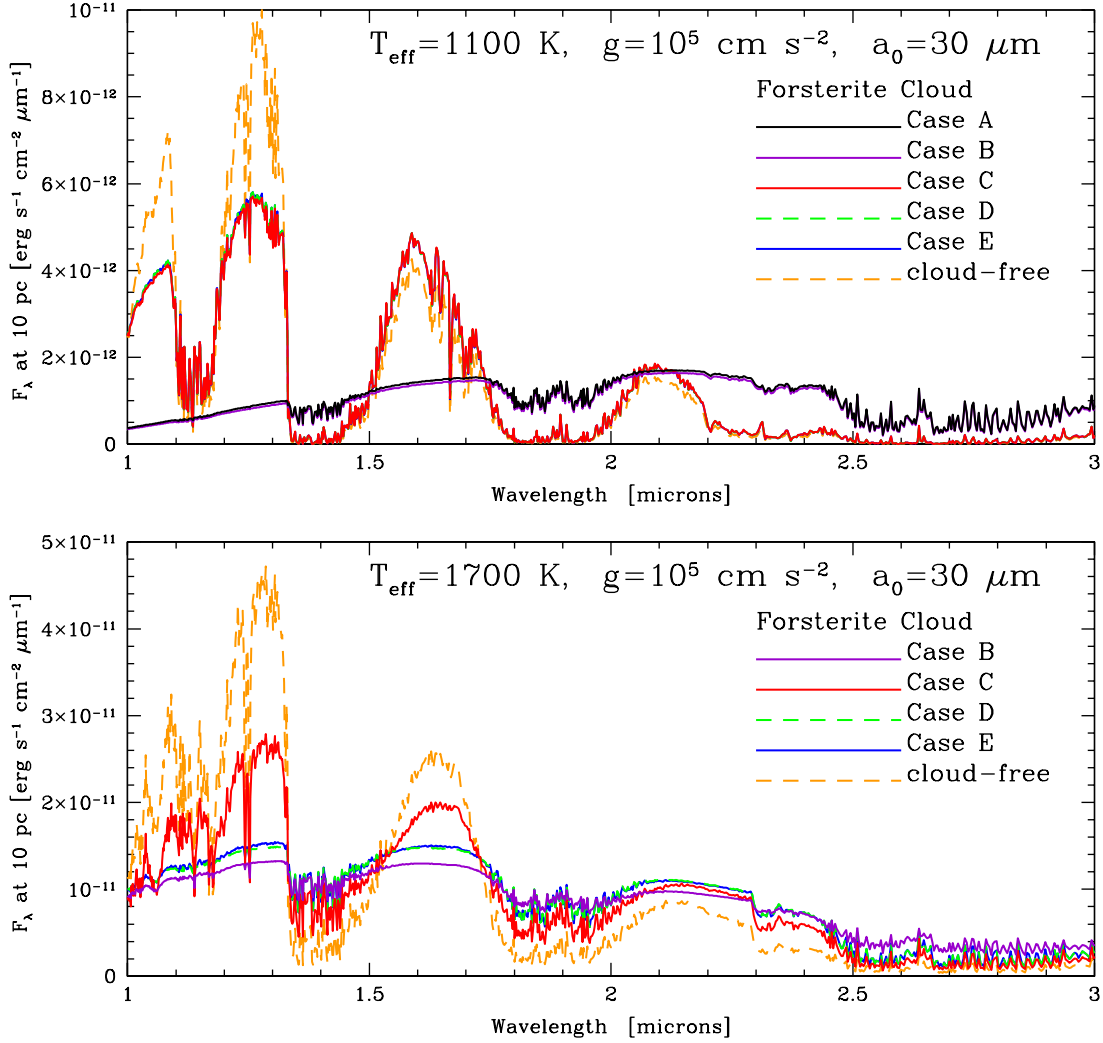}
\caption{\textbf{top panel}: Near-IR spectra of brown dwarfs with
\teff=1100 K and $g$=10$^5$ cm s$^{-2}$
for differing cloud shape prescriptions (A through E).  A forsterite modal
cloud particle size of 30 \mic is assumed.  The spectra produced
using cloud shape models C, D, and E (red, dashed green, and blue
curves, respectively) result in very similar spectra, which largely overlap
on this plot.  In contrast,
cloud shape models A and B produce much redder spectra (black and
purple curves).  For comparison, a cloud-free spectrum for the same
effective temperature is denoted by the dashed orange curve.
\textbf{bottom panel}: Similar to top panel, but for \teff=1700 K.
At this higher \teff, more spectral variation due to cloud shape is
exhibited.  In particular, model C is closer to the cloud-free
case, due to the smaller cloud opacity relative to models D and E.}
\label{fig:2}
\end{figure}

To clarify the role of cloud extent as described in \S\ref{struct}, 
Fig. \ref{fig:2} compares theoretical spectra for
cloud distribution models A, B, C, D, and E at \teffs\ of 1100 and 1700 K,
a gravity ($g$) of 10$^5$ cm s$^{-2}$ and a modal particle radius of 30 \mic.
This figure does not represent our preferred set of models, but rather is
meant to portray the generic differences between the cloud models described in \S\ref{struct}.
At \teff = 1100 K, models C, D, and E are very similar, since the cloud base
where these models differ is all but invisible for this \teff.  These models also 
maintain the large contrast observed for this \teff range between the water
absorption troughs and the classical $Y/Z$, $J$, $H$, and $K$ band peaks.
Models A and B, which allow the cloud to extend 
to the lowest pressures, are severely affected by the cloud opacity, and result
in very flat spectra and a redistribution of flux from short to long wavelengths
that is completely inconsistent with observations (see \S\ref{spectra}).
On the bottom panel of Fig. \ref{fig:2}, we show the corresponding models for
\teff = 1700 K.  Here, the near degeneracy between models C and D/E is broken
and for model C we can discern in the stronger $J$ and $H$ peaks the absence of cloud 
below the forsterite condensation curve.  Clearly, the extent of the cloud(s)
does make a difference.

\begin{figure}
\epsscale{.80}
\plotone{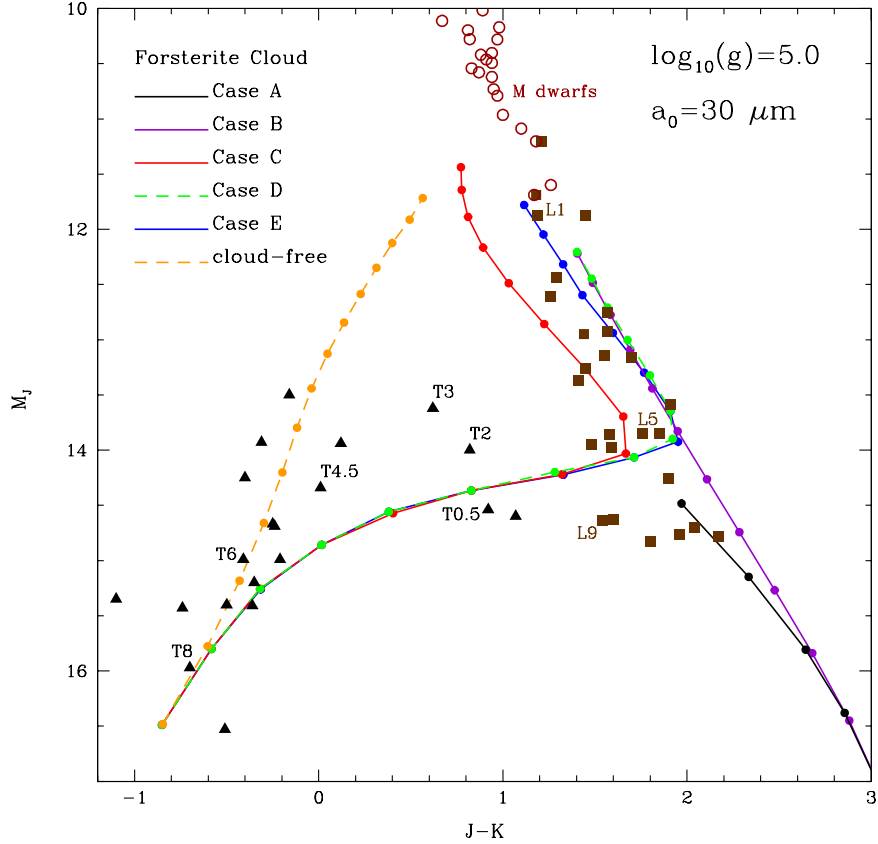}
\caption{Absolute $J$ magnitude versus $J-K$ color for brown dwarf
models with g=10$^5$ cm s$^{-2}$ for 5 different cloud
shape prescriptions and one cloud-free case.  The points along each
model curve are in \teff intervals of 100 K, ranging from 700 K to
1400 K for Case A, 700 K to 2100 K for Cases B, C, and D, and 700 K to
2200 K for Case E and the cloud-free case.
Shown for comparison are L and T dwarf data from Knapp et al. (2004)
and M dwarf data from Leggett (1992).  A forsterite modal cloud particle
size of 30 microns is assumed.  Even in an approximate sense, the
cloud-free model set and model sets A and B do not
come close to fitting the complete set of brown dwarf data in
this color-magnitude space.  On the other hand, model sets
C, D, and E come closer to reproducing the trends seen in the
data, although the brightening in $J$ around the early T dwarfs
is not predicted.  The strong turn toward the blue in $J-K$ with
decreasing \teff near 1400-1500 K is due to the natural deepening of the cloud
position in these model atmospheres. Throughout the paper, we use the MKO photometric system
for the $J$, $H$, and $K$ bands (Tokunaga \& Simons 2002; Tokunaga, Simons, \& Vacca 2003).}
\label{fig:3}
\end{figure}

In Fig. \ref{fig:3}, we compare models A, B, C, D, and E
for 30-\mic particles on an M$_J$ versus $J-K$ color-magnitude (HR) diagram, 
but for a wider range of \teffs than provided in Fig. \ref{fig:2}.  
Included are both the cloud-free (clear) curve and a collection of measured M, L, and T dwarfs
with parallaxes.  All the cloud models (except model A) reproduce the general behavior of the M$\rightarrow$L
transition and the C, D, and E models comfortably inhabit the early-to-mid 
L dwarf regime.  The presence of clouds clearly defines the Ls. In addition, models C, D, and E are degenerate below 
\teff$\sim$1400 K at this gravity and reach the measured mid-T dwarf region to the blue 
in reasonable fashion. Note that the extended cloud models A and B never make the transition to the
``clear" behavior of the late T dwarfs and they are off in $J-K$ by as much as 3 magnitudes at low \teff.  
However, though as Fig. \ref{fig:2} demonstrates the corresponding spectra for
models A and B are disfavored, even these models transverse the L-type spectral
band in reasonable fashion.  The same can be said for the models of Chabrier
et al. (2000) (to which models A and B are closest in their 
particulate spatial distribution) for M$_J$s from $\sim$11 
to $\sim$14.5.  However, depending upon the isochrone chosen, the Chabrier 
et al. models later than $\sim$15.0 magnitudes are redder than models A 
and B by as much as $\sim$0.5 to $\sim$2.0 magnitudes.  This is in part 
a consequence of their use of an ISM particle size distribution.

Furthermore, and importantly, as shown in Fig. \ref{fig:3}, the observations of  
early- to mid-T dwarfs indicate a substantial brightening in $J$, peaking near T3.  
If in fact this observational trend is denoting brightening along an evolutionary track
or along a near-constant gravity/metallicity trajectory, and not an incompletely
sampled scatter in the known population of brown dwarfs with parallaxes 
or the effects of binarity, then models C, D, and E, though they handle 
the early-to-mid Ls and the mid-to-late Ts reasonably well, do not adequately 
reproduce the L$\rightarrow$T transition region.  The diminution of the effects
of cloud optical depth with decreasing \teff is occurring more rapidly in the real objects 
than in the models.  We will discuss this more in \S\ref{profiles} and \S\ref{conclusions}.

\begin{figure}
\epsscale{1.0}
\plotone{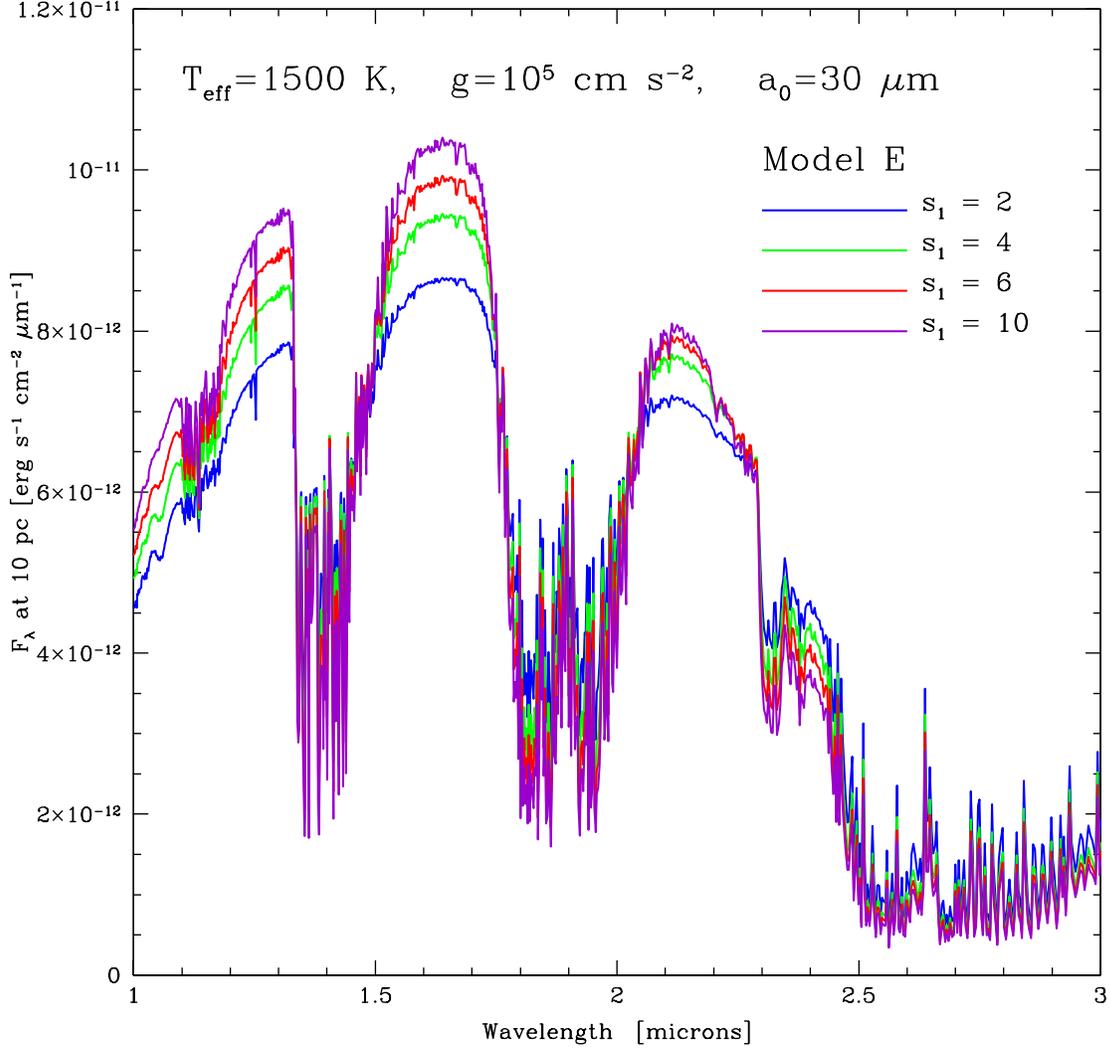}
\caption{Dependence of a model dwarf spectrum on the vertical
extent of the condensate cloud.  A sharper cloudtop cutoff
(larger exponential cutoff parameter, s$_1$) results in a bluer
infrared spectrum.  Our cloud particle distribution model E is assumed.}
\label{fig:4}
\end{figure}

Figure \ref{fig:4} portrays the dependence on the exponential cutoff parameter, 
$s_1$, of the theoretical spectrum of an L dwarf with a \teff\ of 1500 K, a gravity of 10$^{5}$ cm s$^{-2}$, 
a modal particle radius of 30 \mic, and the model E particle density distribution.
The purpose of this diagram is to demonstrate that the upper 
(low-pressure) particle density scale height, a quantity
that is very loosely constrained by first principles and that may be chosen 
by modelers with little physical motivation, makes a difference of as much
as $\sim$20\% in the near infrared fluxes.  Ignorance of this quantity 
can translate into a further ambiguity in an inferred \teff of more 
than 50 K.  The larger the $s_1$ the thinner the upper cloud deck.  
For our subsequent model E spectra, we chose $s_1$ = 6
to reflect our prejudice that the refractory clouds made up of large
particles settle more compactly than terrestrial water clouds (Ackerman \& Marley 2001).

\begin{figure}
\plotone{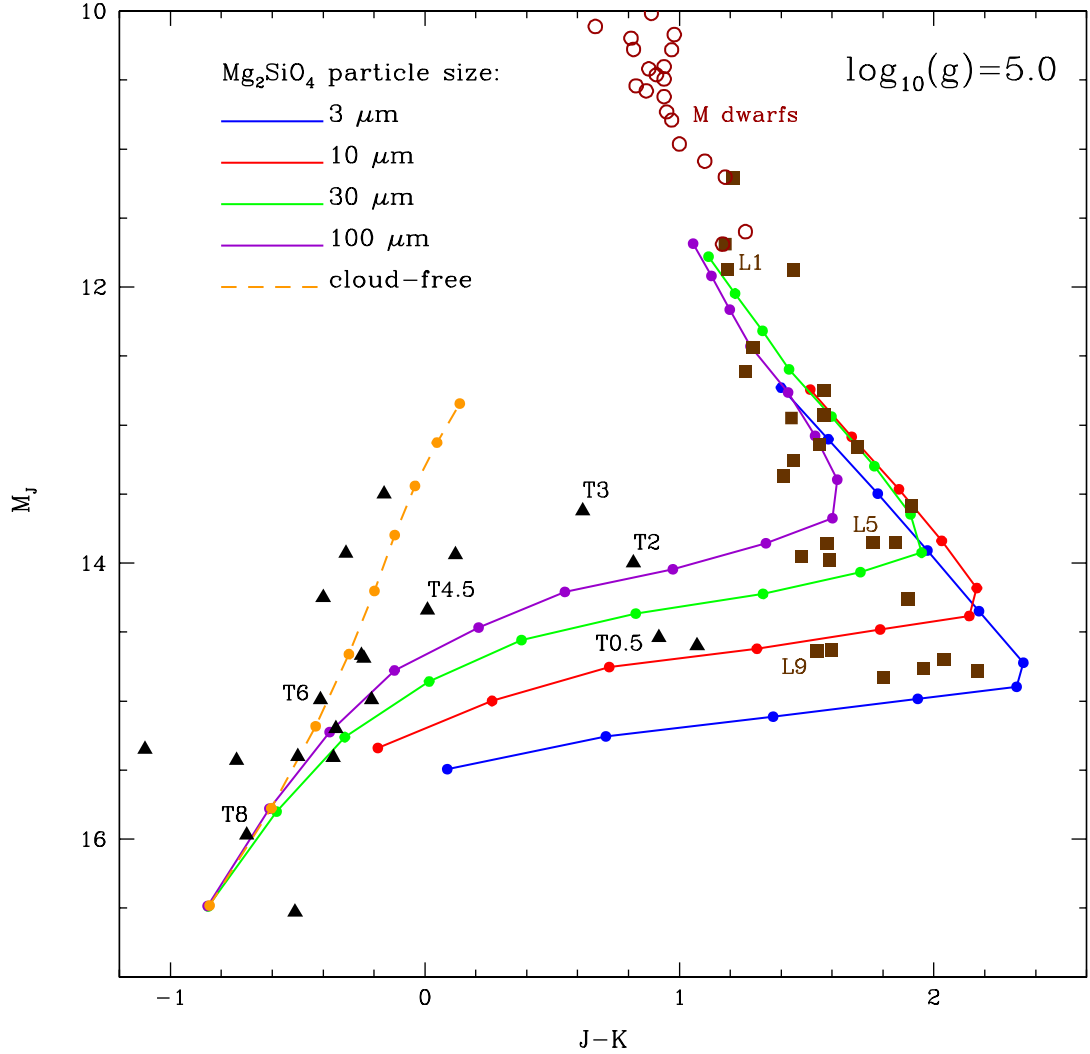}
\caption{Absolute $J$ magnitude versus $J-K$ color for brown dwarf
models with $g$=10$^5$ cm s$^{-2}$.  Four model sets, with modal
forsterite particle sizes of 3, 10, 30, and 100 \mic, along with
an additional cloud-free set, are plotted.  The points along each
model curve are in \teff intervals of 100 K, ranging from 700 K to
1900 K for the 3- and 10-$\mu$m models, 700 K to 2200 K for
the 30- and 100-$\mu$m models, and 700 K to 1500 K for the
cloud-free models.  Shown for comparison are
L and T dwarf data from Knapp et al. (2004)
and M dwarf data from Leggett (1992). At a given \teff, model atmospheres
with larger particles result in stronger $J$-band fluxes.  The strong
turn toward the blue in $J-K$ with decreasing \teff near 1400-1500 K
is due to the natural deepening of the cloud position in these
model atmospheres. The model E cloud shape function is employed.}
\label{fig:5}
\end{figure}

Next, we portray in Fig. \ref{fig:5} an HR diagram similar to Fig. \ref{fig:3},
but as a function of modal grain particle size/radius from 3 to 100 microns.
For comparision, the cloud-free trajectory is included.  \teffs from 700 K
to $\sim$2000 K are represented (see caption) and model E with $s_1$ = 6 is employed.
A constant gravity of 10$^{5}$ cm s$^{-2}$ and forsterite optical properties are assumed.
Data for M, L, and T dwarfs with known parallaxes are superposed.

Figure \ref{fig:5} shows that, at least in M$_J$/$(J-K)$ space, particle size makes 
little difference for the early- to mid-L dwarfs, but a very large difference
in the turnaround elbow and in the M$_J$s for the early- to mid-T dwarfs. The trajectory with 
a larger particle size approaches the anomalous T3 brightness, but fails to include
the latest, reddest L dwarfs.  A small particle size can ``fit" the latest L dwarfs, 
but completely misses the early T dwarfs.  In short, all else being equal
no one particle size and gravity combination can be made to fit all the data,
even from the perspective of this lone HR diagram.  One might conclude from
Fig. \ref{fig:5} that a systematic increase in $a_0$ with decreasing \teff
might be indicated (Knapp et al. 2004), but how this is accomplished by nature would remain
to be determined.  In fact, the simple convective cloud models of
Cooper et al. (2003) and Ackerman \& Marley (2001) predict that particle size slowly decreases
with decreasing \teff.  However, a fast transition to large particles as
\teff decreases, or at some critical \teff, due perhaps to the onset of
runaway particle growth and precipitation in convective flows with large shears
(Shaw 2003; Kostinski \& Shaw 2005), would be worth investigating.

\begin{figure}
\plotone{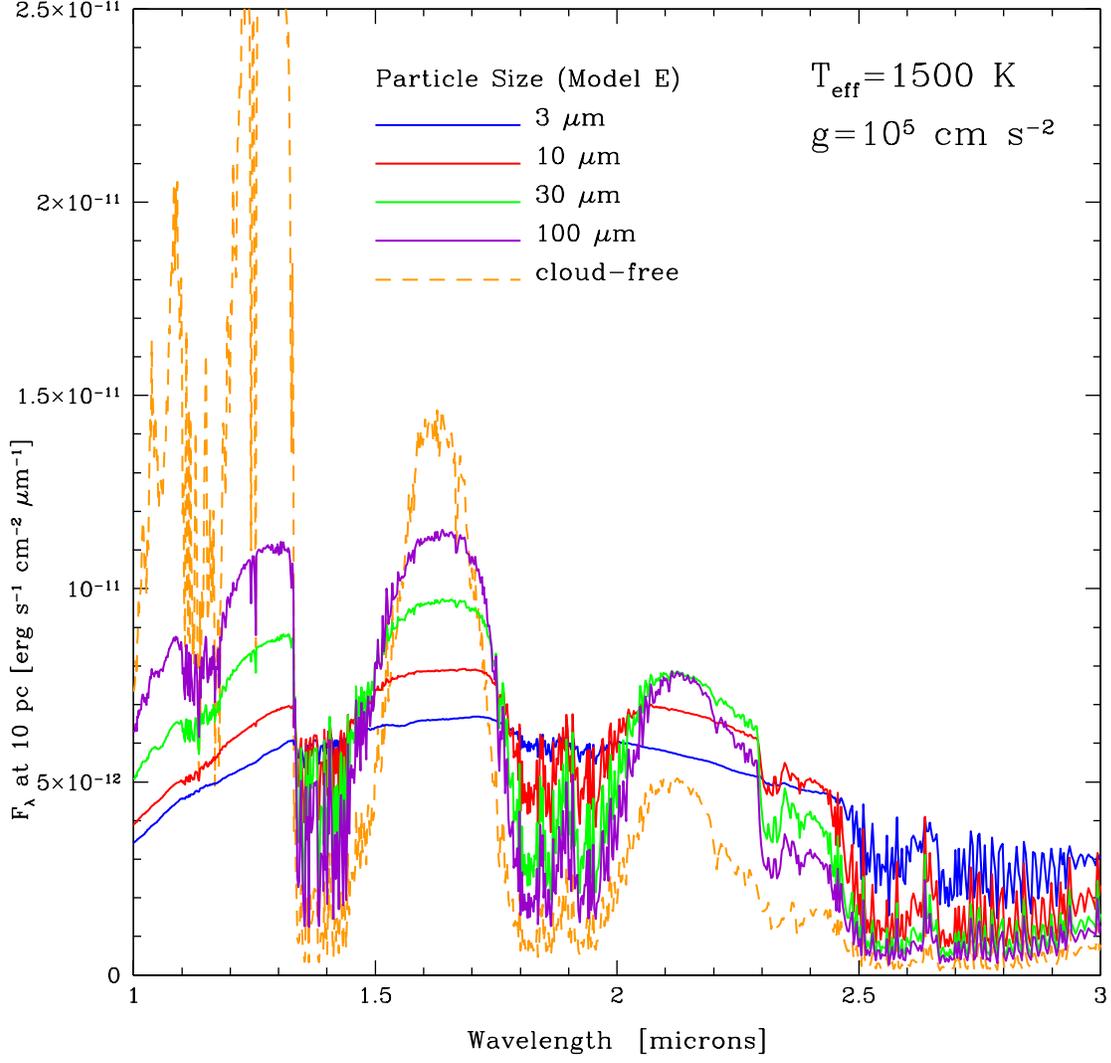}
\caption{Dependence of a model dwarf spectrum on the modal
cloud particle size of forsterite.  Relative to a cloud-free
model (dashed orange curve), near-IR spectral peaks and troughs are
reduced due to the presence of clouds.  Smaller particles result
in redder, smoother spectra. Cloud model E parameters have been assumed.}
\label{fig:6}
\end{figure}

However, as is demonstrated by Fig. \ref{fig:6} for a set of \teff = 1500 K and $g = 10^5$ cm s$^{-2}$
models as a function of modal particle radii and by a basic knowledge 
of L dwarf spectra (as well as of the figures in \S\ref{spectra}), 
small particle sizes below $\sim$30 microns result in theoretical
spectra that are much too smooth, that don't provide the high contrasts across
the water bands generically observed for objects later than mid-L dwarfs, that
suppress the K I doublet in the $J$ band, and that ruin the CO and methane fits
between 2.2 and 2.4 microns.  Hence, small values for $a_0$ are strongly disfavored 
for L and T dwarf models.  The upshot is that while larger particles ($>40$ \mic)
seem favored from general spectral considerations, they make fitting the reddest
late L dwarfs problematic.  From now on, we will take $a_0 = 100$ \mic for
our model E (with $s_1 = 6$) calculations as the best representative parameter
combination and see where that can lead us.  
 
In picking a single particle size for our reference model, we are
not only acknowledging, as we hope others would also do, our current ignorance
of the physics and systematics of clouds in brown dwarf atmospheres,
but our observation that such large particle sizes, whatever the 
grain kinetics, are favored by the data. We find in fact, as 
the following sections show, that our baseline model fits the data rather well,
and certainly no worse than any other model now in the literature.  The reader
is invited to review the dependences on the various cloud parameters we have 
explored in \S\ref{param} to more fully ascertain the limitations of
our baseline model.   

A cloud model must provide the values and systematics with
composition, \teff, and gravity of the modal particle size,
the particle size distribution, the spatial distribution
(with altitude) of cloud properties, the optical constants
of all the relevant grains as a function of wavelength (with 
the effects of grain mantling), the effects of grain asphericity,
the patchiness of the clouds, the effects of convective 
overshoot, and the variations with longitude
and latitude.  Though there have been some good starts at framing 
the discussion of some of these components (Helling et al. 2001, 2004; 
Woitke \& Helling 2003,2004; Ackerman \& Marley 2001; Cooper et al. 2003),  
we believe these issues are still to be adequately addressed. 
Therefore, in this paper we provide our parameter study
and a phenomenological reference model, avoiding what does
not work, while being guided by what seems to, all without
pretending to have achieved an understanding of cloud and
dust physics.

Note that the small abundances
of the calcium/aluminium refractories formed in the radiative zone 
near the M$\rightarrow$L transition, coupled with the higher
per particle opacities of the small particles expected in such zones,
result in a total optical depth that is coincidently comparable to that 
for the 100-micron forsterite particles assumed in our baseline model.
Hence, the total optical depth of large particles of high-abundance 
species (such as forsterite) can crudely mimic the total optical depth
of the smaller/low-abundance particles expected in radiative zones.

\begin{figure}
\plotone{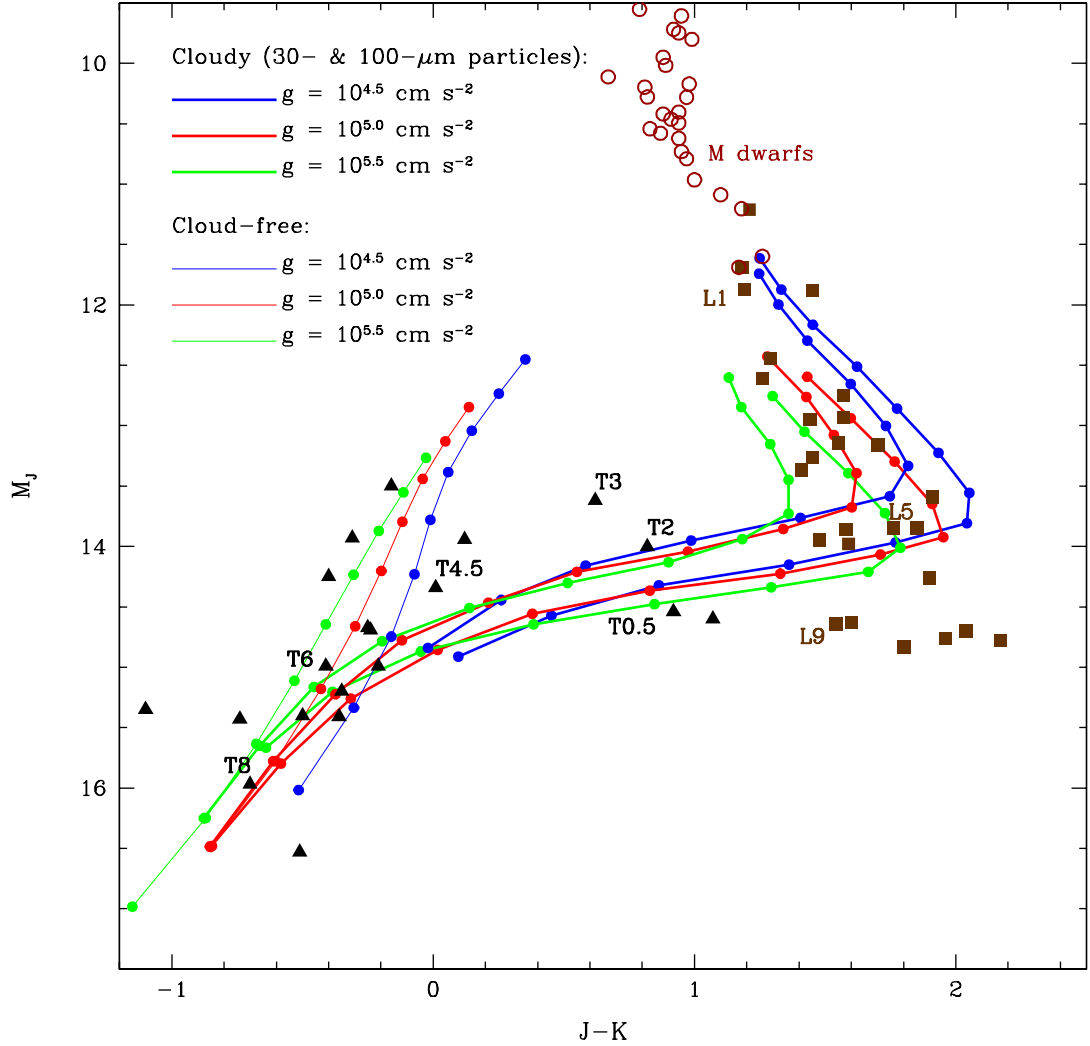}
\caption{Dependence on surface gravity and modal particle
size for models in M$_J$ versus $J-K$ color-magnitude space.
Cloudy (model E) and cloud-free model sets are plotted.  
For each gravity, the larger modal particle size
(100 $\mu$m) falls to the left of the smaller modal size (30 $\mu$m) in
this color-magnitude space.  At a given \teff, a higher gravity
leads to a bluer $J-K$ color and a reduced $J$-band flux.
The points along each model curve are in \teff intervals of 100 K,
ranging from 700 K to 1500 K for the cloud-free models, 900 K to
2100 K for the 30-\mic, $g$=10$^{4.5}$ cm s$^{-2}$ models, 900 K to 2000 K for
the 100-\mic, $g$=10$^{4.5}$ cm s$^{-2}$ models, 700 K to 2200 K
for both cloudy $g$=10$^{5.0}$ cm s$^{-2}$ models, 900 K to 2000 K for the
30-\mic, $g$=10$^{5.5}$ cm s$^{-2}$ models, and 800 K to 2000 K for the
100-\mic, $g$=10$^{5.5}$ cm s$^{-2}$ models.  Shown for comparison are 
L and T dwarf data from Knapp et al. (2004)
and M dwarf data from Leggett (1992).} 
\label{fig:7}
\end{figure}

Before we transition to
a discussion of our fiducial dwarf model set, we present in Fig. \ref{fig:7} 
the dependence on gravity of \teff trajectories in M$_J$ versus $J-K$ space (for $a_0 = 30, 100$ \mic).
Each color represents a different gravity (either 10$^{4.5}$, 10$^{5.0}$, or 10$^{5.5}$ cm s$^{-2}$), 
with the larger gravities to the left.  For each color, the line to the left
is the larger particle size (100 \mic).

The range of gravities employed for Fig. \ref{fig:7} represents a reasonable range of dwarf masses, from 
$\sim$15 \mj to $\sim$70 \mj (Burrows et al. 1997).  That this range of gravities nicely
covers the data points in both the L and T dwarf regimes strongly suggests
that the observed widths of the L and T dwarf bands are explained by gravity
variations, that $g$ is the ``second parameter."  Though one is still 
unsure about the latest Ls and the brightening in $J$ near T3, 
this conclusion seems reasonable and has also been reached  
by Burrows et al. (2002b) and Knapp et al. (2004) using different arguments.
However, it is still possible that binarity can be a factor in the width
of the L and T spectral bands in the HR diagrams.

\section{Temperature/Pressure Profiles}
\label{profiles}

\begin{figure}
\plotone{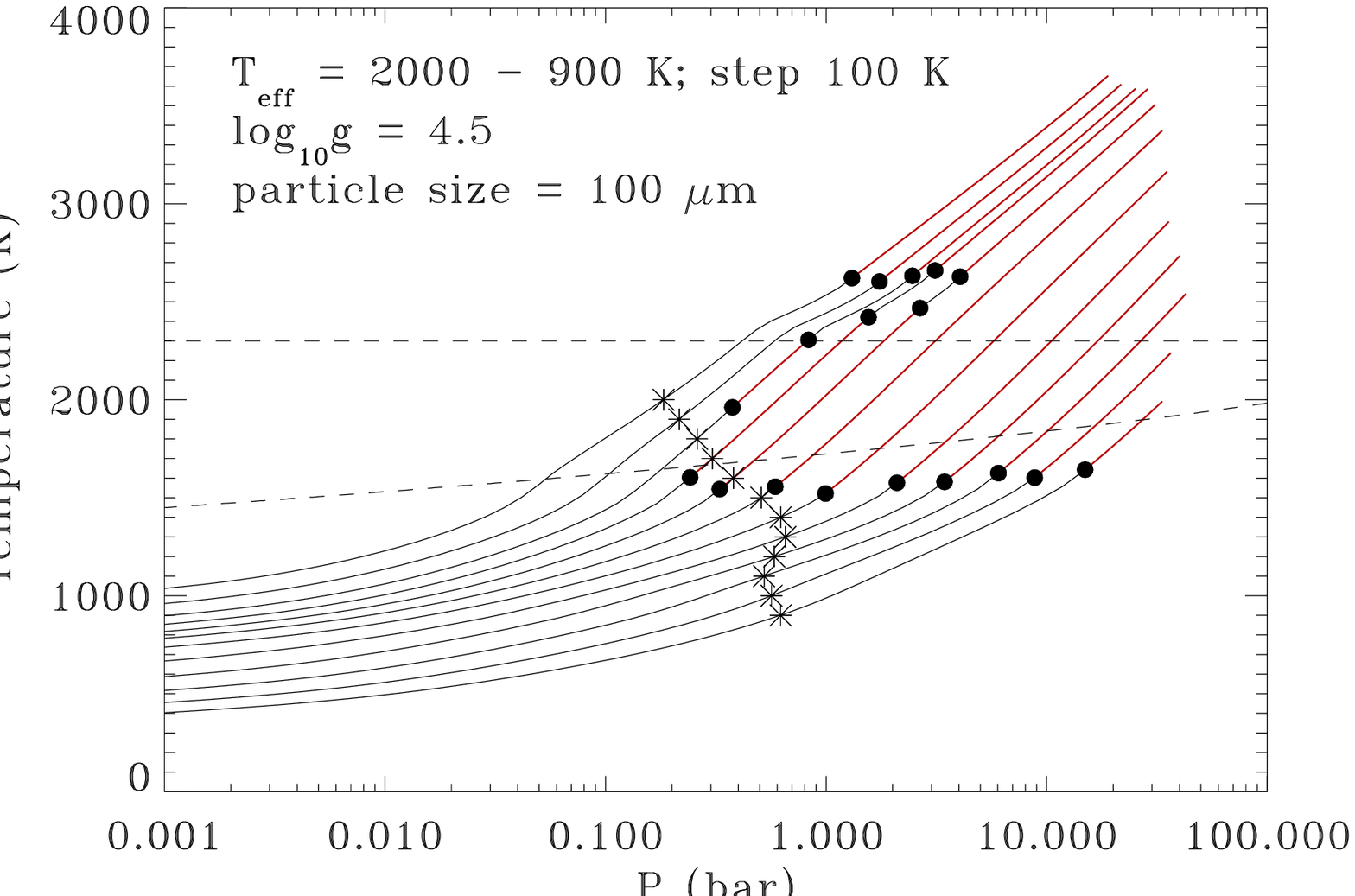}
\caption{Temperature-pressure profiles for a set of models
with constant surface gravity, $\log_{10}g = 4.5$ and
particle size, 100 microns, for different effective temperatures
ranging from \teff = 2000 K (leftmost curve) to 900 K
(rightmost curve). The position where the local temperature is equal
to the effective temperature (which indicates an approximate location
of the region of formation of the emergent radiation) is shown as an asterisk.
The cloud bases are depicted as dashed lines;
the convection zones are drawn in red; and
the black dots show the position of the boundaries of the convection
zone(s). Notice the occurence of two distinct convection zones
for effective temperatures between 1600 K and 1800 K.
}
\label{fig:8}
\end{figure}

\begin{figure}
\plotone{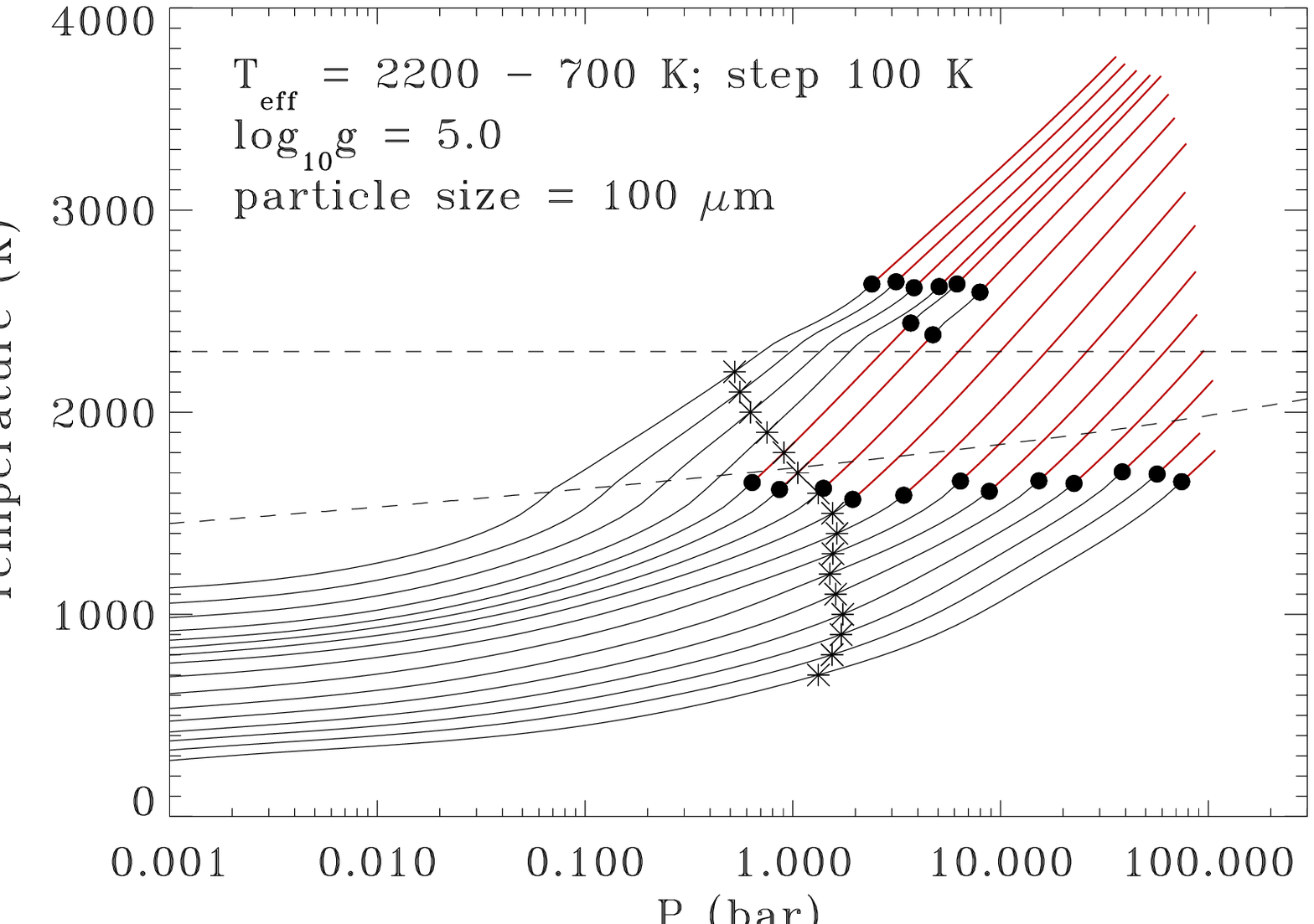}
\caption{Temperature-pressure profiles for a set of models
with constant surface gravity, $\log_{10}g({\rm cm s}^{-2}$) = 5.0 and
particle size, 100 microns, for different effective temperatures
ranging from \teff = 2200 K (leftmost curve) to 700 K
(rightmost curve). The position where the local temperature is equal
to the effective temperature (which indicates an approximate location
of the region of formation of the emergent radiation) is shown as an asterisk.
The cloud bases are depicted as dashed lines;
the convection zones are drawn in red; and
the black dots show the position of the boundaries of the convection
zone(s). Notice the occurence of two distinct convection zones
for \teff  = 1700 K and 1800 K.}

\label{fig:9}
\end{figure}

  Now that we have explored various cloud parametrizations, and settled 
upon one of them to comprise our default set (model E, with $a_0 = 100$ \mic, $s_1 = 6$, and
forsterite's complex index of refraction versus wavelength behavior), let us
return to a discussion of the regions and structure of the generic L or T dwarf
atmosphere.  To do this, we provide in Figs. \ref{fig:8} and \ref{fig:9}
temperature/pressure profiles for different \teffs\ spanning the L 
and T dwarf \teff\ range, at gravities of 10$^{4.5}$ cm s$^{-2}$ and 10$^{5.0}$ cm s$^{-2}$,
respectively.  The red regions are convective, the dots accentuate the convective
boundaries, and the asterisks denote the $T$ = \teff point, roughly the position
of the average photosphere.  The dashed lines indicate the approximate positions of the inner
and outer cloud boundaries, as described in \S\ref{struct}.

As Figs. \ref{fig:8} and \ref{fig:9} show, at \teffs from 2200 K to 1900 K clouds
are radiative and the photospheres are in the radiative zone.  At lower \teffs, from $\sim$1800 K to $\sim$1600 K,
the precise value of which depends upon gravity (and, in fact, particle size and 
metallicity), an isolated outer convective zone emerges that coincides 
approximately with the position of the condensate cloud.  The increasing opacity of the 
particulate matter trips local convection, but at the same time flattens the
T/P profile below this zone at slightly greater pressures.  The upshot is an isolated radiative
zone between two convective zones. Eventually, near $\sim$1600 K for the default model
depicted here, the two convective zones merge.  What happens meteorologically and dynamically
when they merge might make for an interesting study; we will return to this later.  After the two convective zones merge,
the outer convective boundary is always near 1500$-$1600 K and ranges in pressure from
$\sim$0.5 to $\sim$20 bars for the lower gravity and from $\sim$1 to $\sim$80 bars at the higher gravity.
As Figs. \ref{fig:8} and \ref{fig:9} demonstrate, the pressures at the cloud 
base for the two model sets range from $\sim$0.3 to $\sim$20 bars
for $g$=10$^{4.5}$ cm s$^{-2}$ and from $\sim$0.7 to $\sim$100 bars for $g$=10$^5$ cm s$^{-2}$
At a given atmospheric temperature, increasing gravity or decreasing metallicity
increases the local pressure.  Centered at a gravity of 10$^5$ cm s$^{-2}$ and solar metallicity,
at a given $T$ a one order of magntiude range in either gravity or metallicity translates into a variation
of about an order of magnitude in pressure.

The fact that the pressures at the cloud base inexorably rise with decreasing \teff
implies that the optical depth of the cloud increases dramatically with decreasing \teff,
all else being equal.  The Rosseland optical depth of the condensate clouds 
in the models in Figs. \ref{fig:8} and \ref{fig:9} range from of order unity to 
one hundred.  The clouds get thicker with later spectroscopic type from the early
L dwarfs to the late T dwarfs.  

\begin{figure}
\plotone{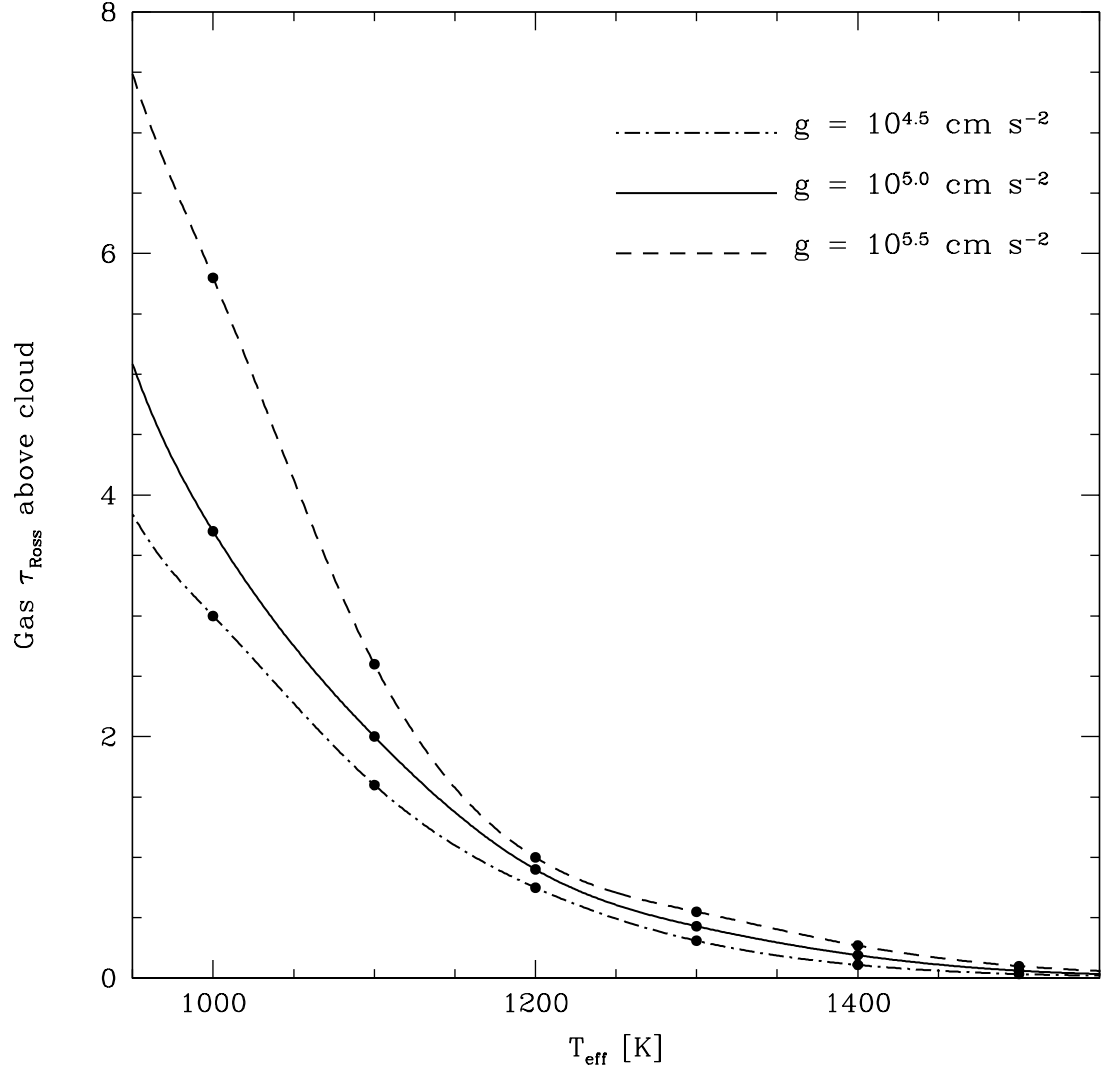}
\caption{Rosseland mean optical depth of gas above cloud versus \teff as
a function of surface gravity.  With decreasing effective temperature,
the clouds become buried beneath an increasing column of
gaseous opacity and so the infrared spectra and colors of these
brown dwarfs approach those of cloud-free atmospheres.}
\label{fig:10}
\end{figure}

However, as the positions of the asterisks indicate, the photosphere ranges   
from $\sim$0.2 to $\sim$0.6 bars at the lower gravity and from $\sim$0.5 to 
$\sim$2.0 bars at the higher gravity.  Its position is tightly constrained to
lie in a narrow pressure range around a bar. As a result, it recedes from the cloud base
and cloud tops with decreasing \teff; as Figs. \ref{fig:8} and 
\ref{fig:9} demonstrate, the top of the convective zone is 
progressively buried and separates from the ``photosphere." This is the origin
of the waning influence as \teff decreases of clouds in dwarf atmospheres and spectra.  
Note that as \teff decreases, the photosphere is first in a 
radiative zone (early L), then the convective zone (mid L),
and finally in the outer radiative zone.  Figure \ref{fig:10} depicts the 
Rosseland depth above the clouds due to gas as a function of \teff for various gravities
and shows how as \teff\ decreases the cloud tops are more and more obscured by 
the cloud-free, gas-dominated regions.

On the basis of the discussion above, one might conclude that the 
L dwarfs, the transition from L to T, and the brightening in the $J$ band depicted
by the data plotted on Figs. \ref{fig:3}, \ref{fig:5}, and \ref{fig:7} all
naturally follow and can be explained by the phenomenology represented in
Figs. \ref{fig:8} and \ref{fig:9}, and most of this is true.  However, the 
L$\rightarrow$T transition appears to occur much more quickly than in our 
fiducial model(s).  The cloud recedes (collapses?) away from the photosphere, or the particle size
increases, faster than is implied by the models of Figs. \ref{fig:8} and \ref{fig:9}, whatever their 
other strengths.  It has been suggested that the clouds break up 
and spawn holes that widen with decreasing \teff (Burgasser et al. 2004). Since the 
clouds inexorably thicken as \teff decreases, we do not favor this break up 
hypothesis.  However, given our current ignorance of cloud meteorology, dynamics, and grain growth, not
the least at the cloud tops that for late-L and early-T dwarfs are closest to the photospheres, we can 
not at present eliminate this scenario.  Be that as it may, we can compare the spectral and color models
we derive with representive observed dwarf spectra to conclude that apart
from the issue of the L$\rightarrow$T transition, our new models reproduce
the observations rather well.  Next, we demonstrate this, using only the coarse
model set in \teff and gravity space we have generated for this paper.

\section{Spectral Comparisons}
\label{spectra}

\begin{figure}
\plotone{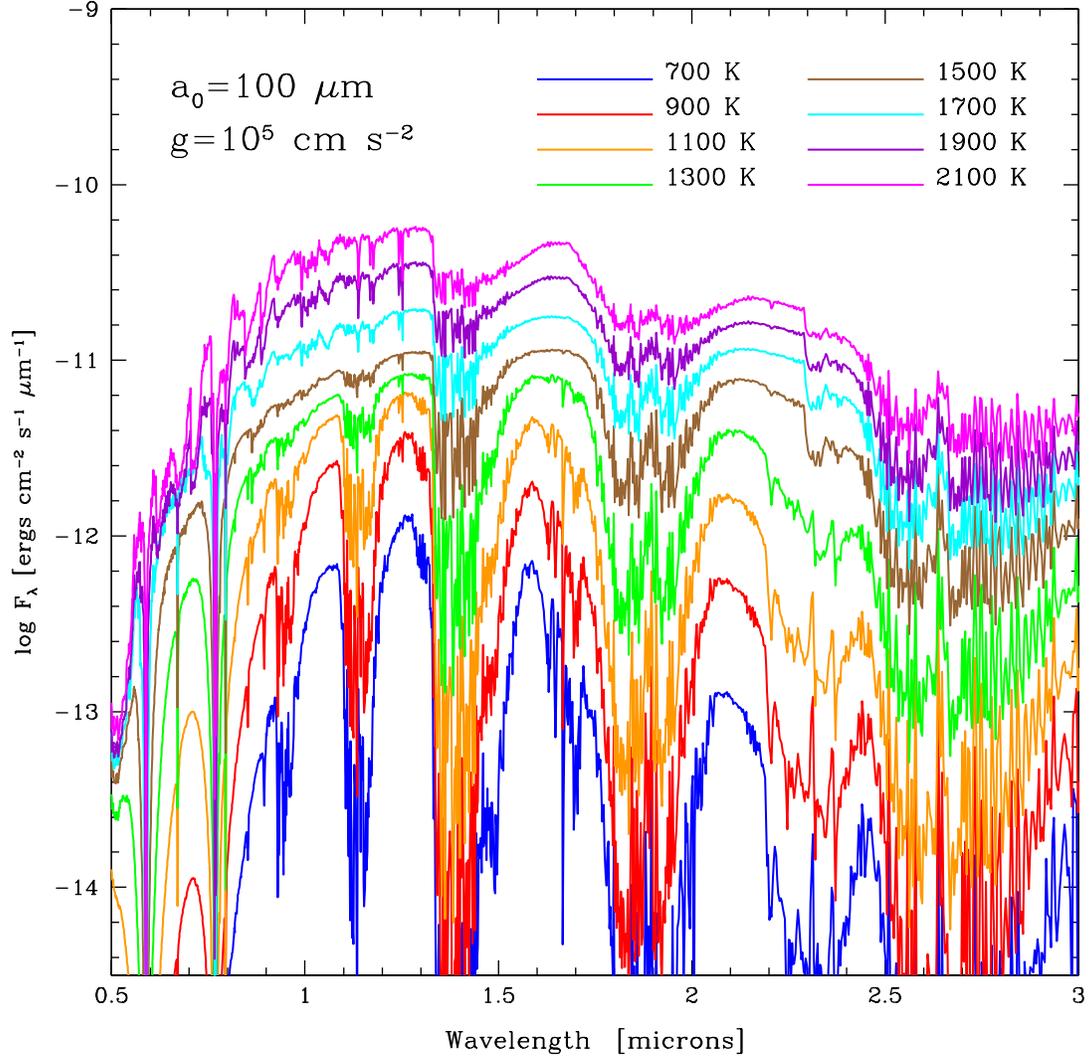}
\caption{A selection of model E spectra (assuming a modal forsterite
particle size of 100 microns) normalized to 10 pc.  Spectra
for eight different effective temperatures from 700 K to 2100 K are shown, encompassing the
range from late T dwarfs to early L dwarfs.  A surface gravity of
10$^5$ cm s$^{-2}$ is assumed and the theoretical radii of Burrows
\etal\ (1997) are used.}
\label{fig:11}
\end{figure}

\begin{figure}
\plotone{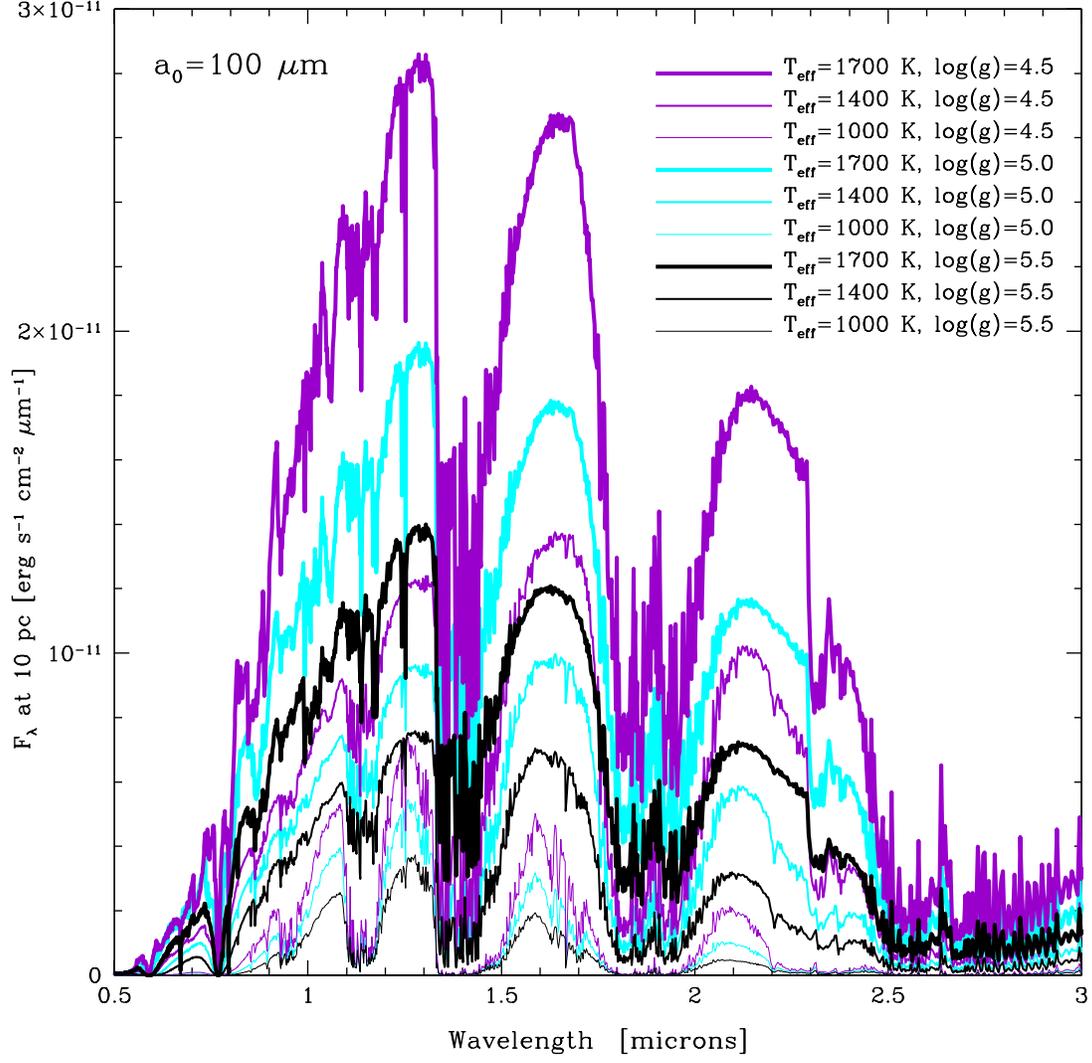}
\caption{Model spectra at effective temperatures
of 1700 K (thick curves), 1400 K (medium curves),
and 1000 K (thin curves) for 3 different surface gravities,
10$^{4.5}$, 10$^5$, and 10$^{5.5}$ cm s$^{-2}$ (purple, cyan, and
black curves, respectively.)  The theoretical brown dwarf radii of Burrows
et al. (1997) are used, and our model spectra are normalized to 10 pc.}
\label{fig:12}
\end{figure}

\begin{figure}
\plotone{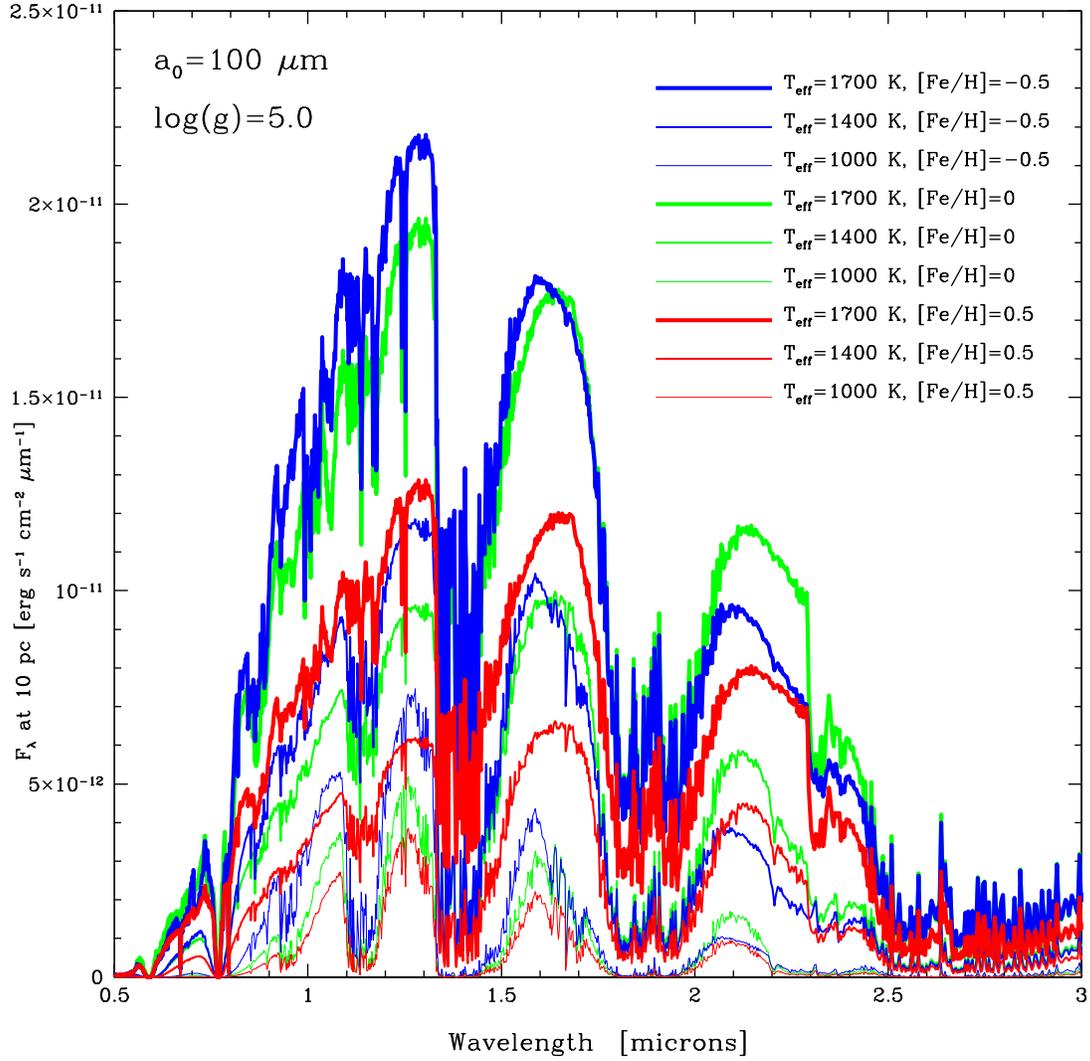}
\caption{Model spectra at effective temperatures
of 1700 K (thick curves), 1400 K (medium curves),
and 1000 K (thin curves) for 3 different metallicities,
[Fe/H]=-0.5, 0, and 0.5 (blue, green, and
red curves, respectively.)  A gravity of 10$^{5}$ cm s$^{-2}$ was employed. 
The theoretical brown dwarf radii of Burrows
et al. (1997) are used, and our model spectra are normalized to 10 pc.}
\label{fig:13}
\end{figure}

Figures \ref{fig:11}, \ref{fig:12}, and \ref{fig:13} summarize the spectra
in the optical and near infrared associated with our baseline model 
set as a function of \teff (Fig. \ref{fig:11}, at solar metallicity), 
gravity (Fig. \ref{fig:12}, for our three gravities), and metallicity 
(Fig. \ref{fig:13}, for three metallicities)\footnote{These spectral 
models are available from the first author upon request.}.  These models fully incorporate an algorithm 
for condensation clouds and the latest gas-phase opacities.  As Fig. \ref{fig:13} indicates,
and as one would generally expect, lower metallicity models are bluer
(compare the green, red, and blue curves at the $Y$/$Z$, $J$, and $H$ peaks).  However, some of the 
non-monotonic behavior in, for instance, the $K$ band can be explained by the fact that 
the CO/CH$_4$ ratio, the atmospheric pressures, and the contribution 
of clouds are all metallicity-dependent, sometimes in countervailing ways. 

In addition to incorporating the evolving effects of grains as \teff decreases, these 
spectral models track the transition from CO to CH$_4$, particularly manifest in the $K$ band in 
the 2.3 \mic to 2.2 \mic switch, the deepening of the methane feature in the $H$ band, 
the appearance (then disappearance) of FeH and CrH features near $\sim$1.0 \mic
and $\sim$0.85 \mic, the emergence of the neutral Na, K, Cs, and Rb alkali features 
below 0.9 \mic, the emergence of the $Y$/$Z$ band peak (near 1.05 \mic), the disappearance
of TiO and VO, and the resculpting of the $J$, $H$, and $K$ bands.  The shape change
in the $K$ band due in part to collision-induced H$_2$ absorption is particularly striking.
Also seen is the evolution of the K I features in the $J$ band near 1.25 \mic and near 1.17 \mic.

\begin{figure}
\plotone{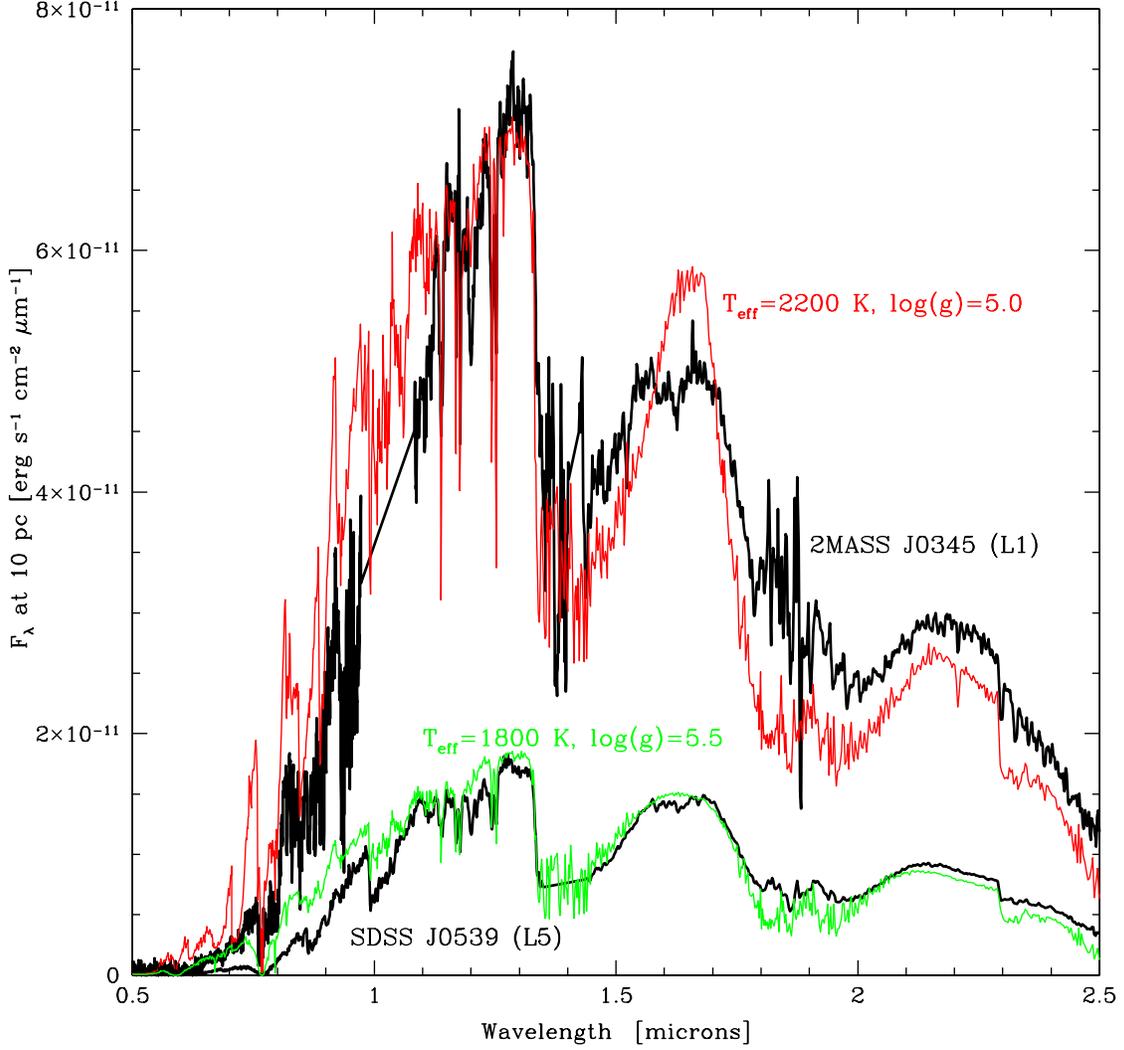}
\caption{Spectrum of 2MASS J03454316+2540233 (Leggett \etal\ 2001;
Kirkpatrick \etal\ 1997), an L1 dwarf, and SDSS J05395199-0059020  
(Geballe \etal\ 2002; Leggett \etal\ 2000), an L5 dwarf.
Both spectra have been normalized to a their absolute
fluxes at 10 pc.  Shown for comparison are model spectra,
also normalized to 10 pc, from our coarse model grid, and
using the theoretical radii of Burrows \etal\ (1997).
These are by no means model spectral fits to the data.  Rather, they
are absolute comparisons between observational and
theoretical results. Model E parameters are assumed.}
\label{fig:14}
\end{figure}

\begin{figure}
\plotone{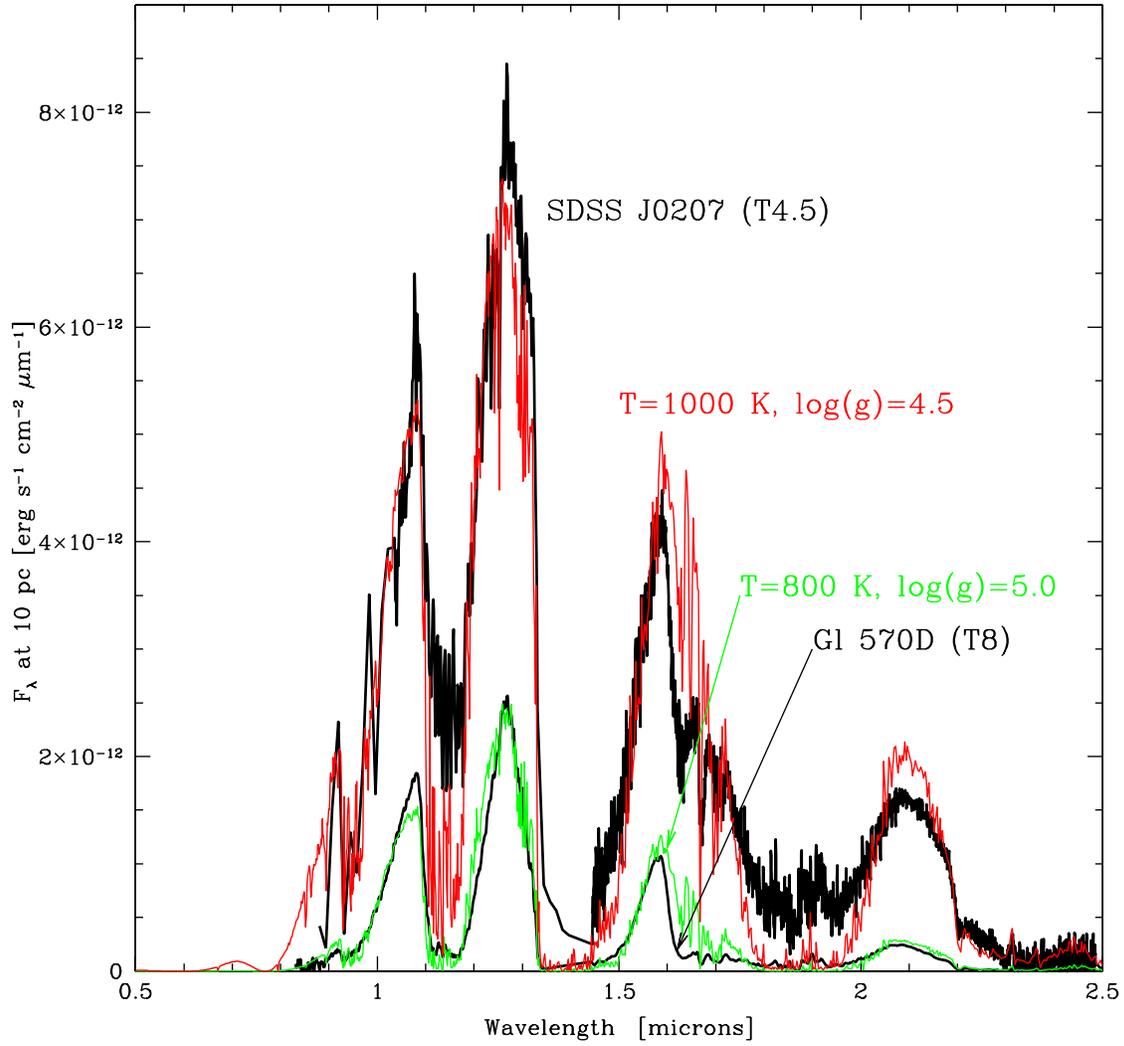}
\caption{Spectrum of SDSS J02074248+0000562 (Geballe \etal\ 2002;
Tsvetanov \etal\ 2000), a T4.5 dwarf, and Gl 570D
(Geballe \etal\ 2001), a T8 dwarf.  As in Fig. \ref{fig:14},
the spectra have been normalized to 10 pc for an absolute
comparison with theoretical model spectra.}
\label{fig:15}
\end{figure}

Without attempting detailed fits, we compare in Figs. \ref{fig:14} and \ref{fig:15}, 
the closest representatives of our model set at solar metallicity with four observed spectra for the L1 dwarf
2MASS J03454316+2540233 (Kirkpatrick et al. 1997; Leggett et al. 2001), the L5 dwarf SDSS J05395199-0059020 
(Leggett et al. 2000; Geballe et al. 2002), the T4.5 dwarf SDSS J02074248+0000562 (Tsvetanov et al. 2000;
Geballe et al. 2002), and the T8 dwarf Gliese 570D (Burgasser et al. 2000; Geballe et al. 2001).
All the measured spectra are for objects with measured parallaxes (Vrba et al. 2004) and 
Figs. \ref{fig:14} and \ref{fig:15} are {\it absolute}, not relative,
comparisons. Had we generated a new set of models with a finer gravity, \teff, and metallicity
grid, we could have obtained even better fits, but have left that exercise to a future paper.

For the L1 dwarf 2MASS J0345, though the $H$-band comparison is slightly problematic, the
correspondence between theory and observation in the shape of the spectrum from 1.8 to 2.3 \mic 
and in the $J$ band is quite good.  For the L5 dwarf SDSS J0539, the correspondence throughout
the spectrum depicted is excellent, and even the FeH Wing-Ford feature at $\sim$0.99 \mic is
reasonably well reproduced.  Except for shape discrepancies in the $H$ and $K$ bands, the correspondence
between theory and observation from 0.9 \mic to 1.6 \mic for the T4.5 SDSS J0207 is striking.
Furthermore, for the T8 dwarf Gliese 570D, except for deviations at the $Y/Z$ peak and in the 
$H$ band (the latter due, again, to problems with the extant methane database), the ``fit" with theory
is remarkable.  Note that all these comparisons are being made for spectra plotted linearly, not
in the logarithm, and are, therefore, that much more encouraging.  

The successes for the L dwarf comparisons seen in Fig. \ref{fig:14} suggest that
for the early to mid-L dwarfs our condensation cloud model is rather good, surprisingly so
given the possible variations on cloud modeling.  Furthermore, Fig. \ref{fig:15} suggests that
for the mid- to late T dwarfs the models are also good.  For all four representative dwarfs depicted,
the \teffs and gravities shown on the plots are useful first guesses for these
quantities. 
As Marley, Cushing, \& Saumon (2004) suggest, for T dwarfs later than about T5 we find
that the clear models without any direct effects of clouds fit the observations
in the near to mid-infrared slightly better than the cloudy models.  Marley, Cushing, \& Saumon (2004)
claim that this demarcation is near \teffs of 1200 K.  However, we determine that the 
effects of clouds, particularly from the optical to $\sim$1.1 microns (Burrows et al. 2002),
completely disappear at \teffs cooler than $\sim$1100 K.   This slight difference highlights 
the problems that attend the L$\rightarrow$T transition.  Beyond \teffs of 1100 K we discern
little difference at any wavelength for which comparisions can be made between 
the Marley, Cushing, \& Saumon (2004) models and the models presented in this paper.

\section{Color-Magnitude Results}
\label{col-mag}

\begin{figure}
\plotone{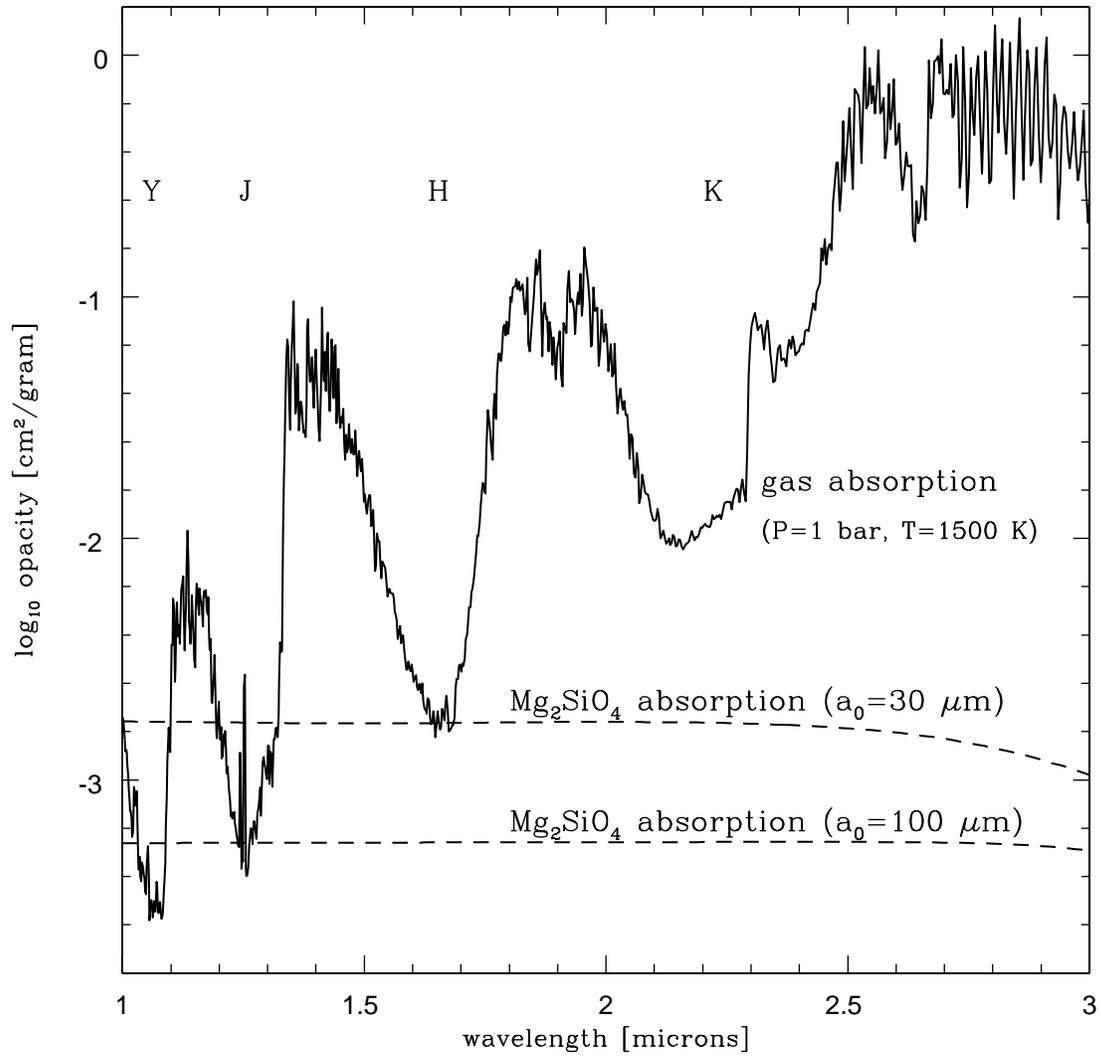}
\caption{Abundance-weighted comparison of forsterite opacity
for 30-\mic and 100-\mic modal particle sizes with that of the
total gas opacity at a pressure of 1 bar and temperature of
1500 K.  In the $Y$/$Z$ and $J$ bands, forsterite can be a dominant
opacity source, depending upon the depth of the cloud layer
in the atmosphere.}
\label{fig:16}
\end{figure}

If there were to be a brightening due to the diminution of the effects of clouds,
it is not surprising that this would be most manifest in the $J$ and $Y/Z$ bands.
Figure \ref{fig:16} provides a comparison between the forsterite and the solar-metallicity gas opacities
for two particle sizes and a typical atmospheric thermodynamic point.  The gas-phase
opacities show the classical variation due to the water absorption troughs that define
the near infrared photometric bands.  At $\sim$1.05 \mic and $\sim$1.25 \mic,
the grain opacities are most competitive with the gas, indicating clearly that it is 
in the $Y/Z$ and $J$ bands that the recession to depth, thinning, or disappearance of the clouds would
first be obvious.  Hence, it is all the more perplexing that our default models
can not reproduce the $\sim$1.0 magnitude rise seen in M$_J$ 
versus $J-K$ near subtype T3, if ``rise" it is.  Nevertheless, it is instructive
to see just what the new models do predict.  Figure \ref{fig:17} is a M$_J$ versus $J-K$
HR diagram with \teff trajectories for all our fiducial models for the three gravities
(10$^{4.5}$, 10$^{5.0}$, and 10$^{5.5}$ cm s$^{-2}$) and three metallicities ([Fe/H] = \{-0.5, 0, +0.5\}).
The cloud-free models and the measured points for M, L, and T dwarfs with parallaxes are also included on the plot.
A given color denotes a given metallicity, the leftmost models being the most metal-poor.
The green curves are for solar metallicity.
For a given triplet, the leftmost model is the highest gravity. 

Consonant with the discussion in \S\ref{param}, the early to mid-L dwarfs and the mid- to late-T dwarfs
fit quite well.  The thicknesses of both the T and L dwarf bands are well recreated by
variation in gravity, with perhaps a slight metallicity dependence.  The gravity dependence
in the L$\rightarrow$T transition is expected to be weak, while the metallicity dependence
is strong, as large as could be produced by a variation in particle size of a factor of ten.
The latest L dwarfs are still outliers in the theory, for all \teffs, gravities, and metallicities.
It is indeed ironic that to put models among these latest L dwarfs requires higher cloud
opacities, while to explain the anomaly near T3 requires lower cloud opacities, emphasizing
that the swing from L9 to T3 is very rapid, even more rapid that the discussion and plots
in \S\ref{profiles} provide for.

As Fig. \ref{fig:17} indicates, we could fit the $J$ ``brightening" around T3 with 
lower metallicities, only a factor of $\sim$3 below solar.  However, 
as Fig. \ref{fig:17} also indicates, such a metallicity,
though it might help with some T dwarf fits, is not consistent with most of them.
Furthermore, if the observed trajectory from L to T is not very incomplete,
then the metallicity explanation would be hard-pressed to explain the empty regions
in M$_J$/$(J-K)$ space that should be occupied by low-metallicity dwarfs.

The corresponding diagram for M$_K$ versus $J-K$ (Fig. \ref{fig:18}) is more encouraging, with the 
green solar-metallicity lines inhabiting the L dwarf, the L$\rightarrow$T transition,
and the T dwarf regions rather well.  However, the latest L dwarfs are still
left out. Note that Figs. \ref{fig:17} and \ref{fig:18} suggest 
that the low-metallicity ([Fe/H]$\le$-1.0) subdwarf L dwarfs will swing to
the blue earlier than solar-metallicity L dwarfs and, hence, that the extent
of the subdwarf L dwarf family in color and \teff space will be smaller.  At low enough
metallicity, the L dwarfs might altogether disappear as a class. 

\begin{figure}
\epsscale{.80}
\plotone{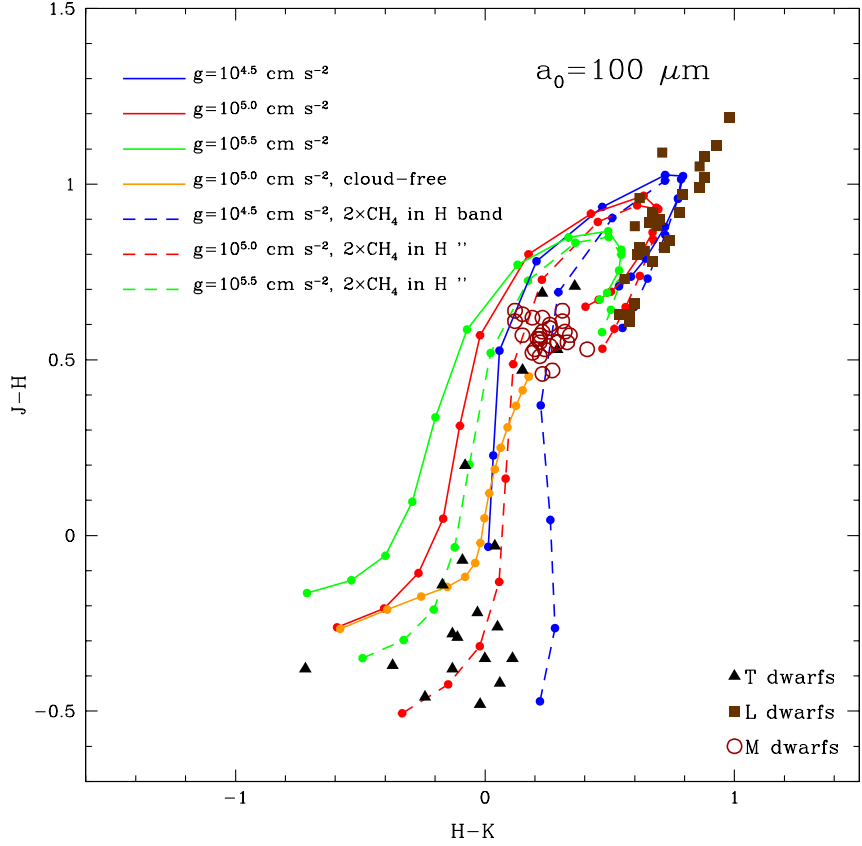}
\caption{$J-H$ versus $H-K$ color-color diagram.  Assuming a
modal forsterite particle size of 100 \mic, models sets
for surface gravities of 10$^{4.5}$, 10$^{5.0}$, and 10$^{5.5}$
cm s$^{-2}$ are plotted (blue, red, and green solid curves, respectively).
In addition, we have plotted a cloud-free model set for
g = 10$^{5.0}$ cm s$^{-2}$ (orange curve).  Because our $H$-band opacity database is
missing the ``hot bands'' of methane, we estimate the effects
of the expected additional opacity by artificially enhancing
the methane opacity in the $H$-band by a factor of 2 (blue, red,
and green dashed curves). The points along each model curve are
in \teff intervals of 100 K, ranging from 900 K to 2000 K for
the fiducial $g$=10$^{4.5}$ cm s$^{-2}$ model, 700 K to 2200 K for the
fiducial $g$=10$^{5.0}$ cm s$^{-2}$ model, 800 K to 2000 K for the fiducial $g$=10$^{5.5}$ cm s$^{-2}$
model, 800 K to 2000 K for the enhanced $g$=10$^{4.5}$ cm s$^{-2}$ and $g$=10$^{5.5}$ cm s$^{-2}$
models, 700 K to 2100 K for the enhanced $g$=10$^{5.0}$ cm s$^{-2}$ model,
and 700 K to 2100 K for the cloud-free model.  Shown for comparison are
L and T dwarf data from Knapp et al. (2004) and M dwarf data from
Leggett (1992). A model E cloud shape function is employed.}
\label{fig:19}
\end{figure}

Figure \ref{fig:19} depicts the $J-H$ versus $H-K$ color-color diagram for
a subset of the new models and explains a curious behavior.  As before, each line
is for a given gravity and all models are for solar metallicity.  Data for T, L, and M dwarfs and
a cloud-free trajectory (orange) are also shown.  As the data indicate, there is a distinct clump
of M dwarfs.  To the red in both colors can be found the L dwarfs, also as a clump.
However, the T dwarfs are found on the opposite side from the L dwarfs, re-traversing
the M dwarf realm and extending to the blue.  The clear models do not extend redward of
the M dwarfs.   However, the cloudy models do show this behavior, curling
up to the L dwarfs, then back through or near the M dwarfs, and on to the blue T dwarfs.
This sinus behavior is a consequence of the appearance, then disappearance of the clouds
from the photosphere.  The best fit is for models (the dashed curves) for which the
CH$_4$ opacity in the $H$ band has been artifically increased by a factor of 2 to account for
the missing CH$_4$ hot bands.  Our default model set does not incorporate this artifical
enhancement, but we present such models here to show the better fits that are 
possible with the same opacity increase as is required for better spectral fits
to measured T dwarfs in the $H$ band. 

\begin{figure}\epsscale{1.0}
\plotone{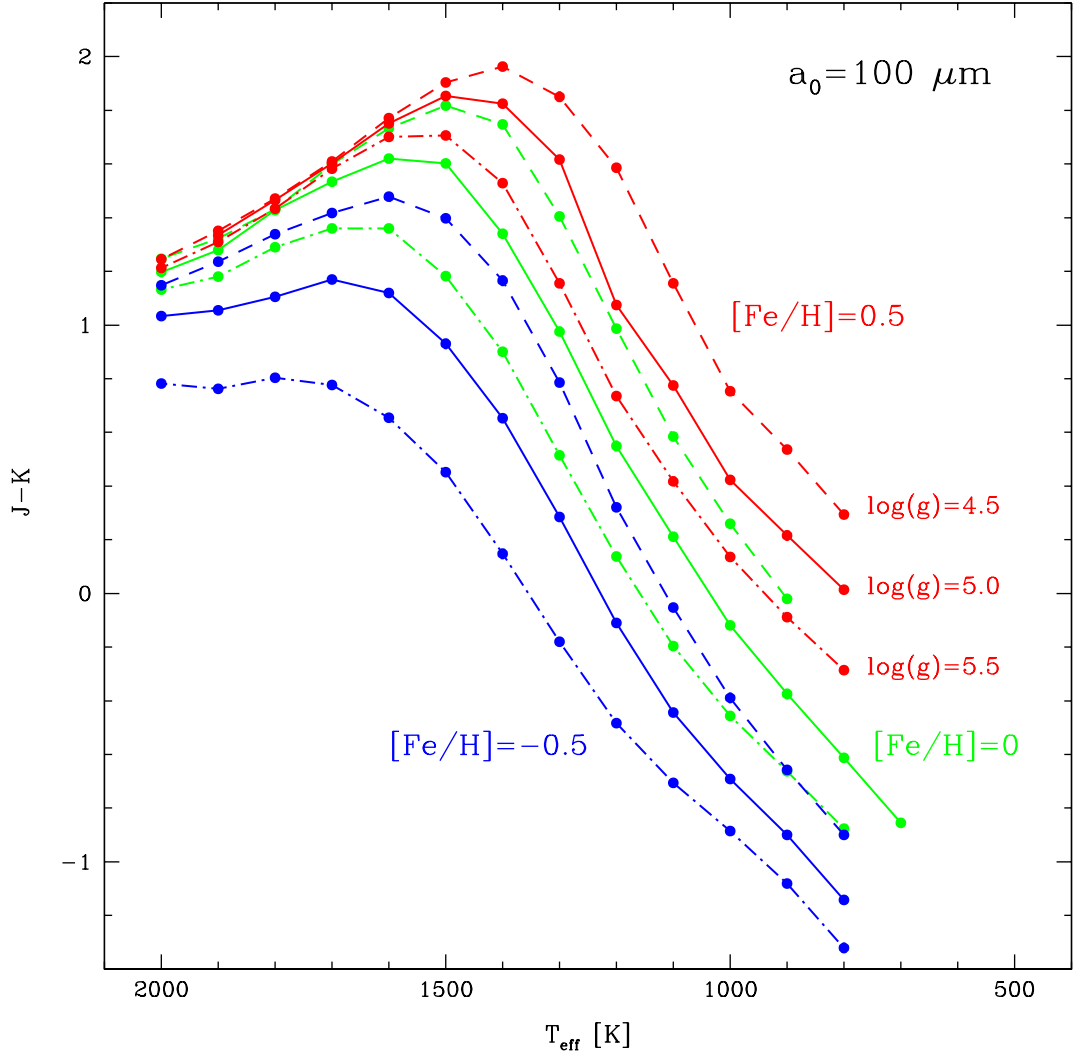}
\caption{Metallicity and surface gravity dependence of model E dwarf models in $J-K$
versus \teff space.  The $J-K$ color tends to peak at its reddest value in the
1400-1600 K \teff range. $J-K$ varies significantly with both
metallicity and gravity, but it clearly turns bluer at cooler
effective temperatures.  The bluest colors are exhibited for low
\teff, low metallicity, and high surface gravity.}
\label{fig:20}
\end{figure}

Figure \ref{fig:20} is a summary diagram providing $J-K$ versus \teff trajectories for a variety of
metallicities and gravities.  The basic behaviors of the entire M, L, and T dwarf families
are encapsulated here, though the slope from the late L dwarf bump to the T dwarfs
at lower \teffs is too shallow to explain the anomalous brightening in $J$.

\begin{figure}
\plotone{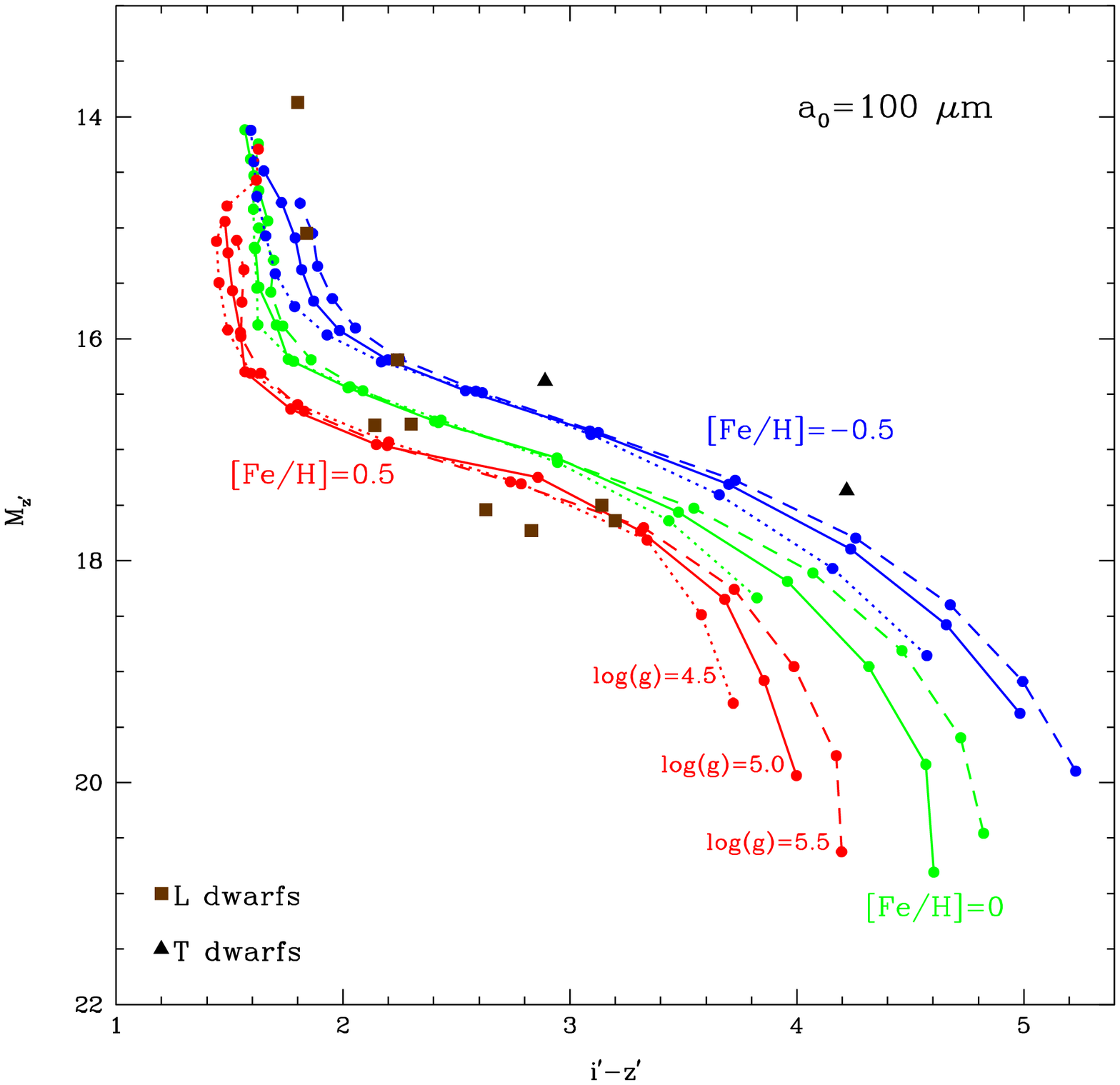}
\caption{Metallicity and surface gravity dependence of dwarf models in absolute
$z^{\prime}$ magnitude versus $i^{\prime}-z^{\prime}$ color space.  Sloan filter
functions and zero points were used.  Cloudy models
(model E) with a forsterite modal particle size of 100 \mic
are shown for surface gravities of 10$^{4.5}$, 10$^{5.0}$, and
10$^{5.5}$ cm s$^{-2}$ (dotted, solid, and dashed curves,
respectively), and for metallicities, [Fe/H] = -0.5, 0, and 0.5 (blue, green, and
red curves, respectively). Shown for comparison are
L and T dwarf data from Knapp et al. (2004).}
\label{fig:21}
\end{figure}

Finally, in Fig. \ref{fig:21} we portray model trajectories for the various metallicities 
and gravities of our new model set in the M$_{z^{\prime}}$ versus i$^{\prime}$-z$^{\prime}$
color-magnitude diagram\footnote{z$^{\prime}$ and i$^{\prime}$ are calculated using 
the Sloan filters.}.  L and T dwarf data from Knapp et al. (2004) and for which
there are parallaxes are superposed.  The theory makes a passable job at following the
dwarf points, though for the few T dwarfs included a better fit is indeed with
the lower metallicity lines, while for the latter L dwarfs a slightly higher metallicity 
would be preferred.  More data in these bands for dwarfs with parallaxes is clearly needed. 

Figures \ref{fig:11} through \ref{fig:21} represent a new generation of models and
the degree to which they reproduce current L and T dwarf data.  All in all, the new models are
an advance and they can be used with profit to extract physical information from
observed spectra, particularly for the early- to mid-L dwarfs and the mid- to late T dwarfs.  
However, a number of questions remain unresolved, in particular the character of the $J$
band brightening and reason for the redness and dimness of the latest L dwarfs.

\section{Conclusions and Discussion}
\label{conclusions}

Using an updated gaseous opacity database and a model for clouds in L and T dwarf
atmospheres, we have generated and presented a new series of spectral models
spanning the \teff\ range from 2200 K to 700 K, as a function of gravity and metallicity.
The correspondence with observed spectra and infrared colors for early- and mid-L dwarfs 
and for mid- to late-T dwarfs is good.  We find that the width of both the T and L dwarf branches
is naturally explained by reasonable variations in gravity and, therefore, that gravity is 
indeed the ``second parameter" of the L/T dwarf sequence, \teff being the first.  We have explored the dependence
of theoretical dwarf spectra and HR diagrams upon various cloud parametrizations, such 
as particle size, cloud scale heights, and spatial distribution.  The results are a cautionary
tale that our current ignorance of detailed cloud meteorology renders ambiguous the extraction
of various physical quantities such as \teff and gravity for mid-L to early-T dwarfs. 
We estimate that errors of $\sim$50$-$100 K in \teff\ and $\sim$0.3 in $\log_{10} g$ are
likely, particularly near \teff$\sim$1100$-$1500 K.  Nevertheless, 
for decreasing \teff, we capture with some accuracy the transition from CO 
to CH$_4$ in the $K$ band, the FeH and CrH features at $\sim$1.0 \mic
and $\sim$0.85 \mic, the emergence of the neutral Na, K, Cs, and Rb alkali features,
the growth of the $Y$/$Z$ band peak, the disappearance of TiO and VO in L dwarfs, 
the transformations of the $J$, $H$, and $K$ bands, and the 
evolution of the K I features near 1.25 \mic and near 1.17 \mic. Moreover, we
speculate that the subdwarf branch of the L dwarfs will be narrow, and, that for low enough
metallicity, the L dwarfs will disappear altogether as a spectroscopic class.  Furthermore,
we note that the new, lower solar oxygen abundances of Allende-Prieto, Lambert, \& Asplund (2002) 
produce better fits to brown dwarf data than do the older Anders \& Grevesse (1989) values. 
However, we do not reproduce the ``brightening" in $J$, nor the 
dimness of the latest L dwarfs. 

What could be the reason for the problems near T3 and L8/9?  The $J$ and $Y/Z$ band brightening
will require a more rapid decrease in atmospheric cloud opacity than occurs naturally
in all the uniform cloud models published to date, including our own.  Those of Tsuji,
Nakajima, \& Yanagisawa (2004) (their ``UCM" set) do not reproduce either anomaly 
-- neither do the models of Ackerman \& Marley (2001), Marley et al. 
(2002), Chabrier et al. (2000), or Allard et al. (2001). Burgasser et al. 
(2004) have postulated the break up of the clouds near a \teff of 1200$-$1300 K
and the appearance of holes, whose filling fraction increases across the L$\rightarrow$T transition
until that fraction is unity.  A virtue of this model is the natural explanation
of the apparent resurgence of the FeH features in the early- to mid-T dwarfs 
(Burgasser et al. 2004; McLean et al. 2003; Cushing, Rayner, \& Vacca 2005).
The FeH abundances near the photospheres should be waning; holes
could allow us to see more deeply to the higher-temperature regions in which the FeH abundance is 
large.  The break-up model might imply larger temporal variations in the FeH, $J$, and $Y/Z$
band fluxes for early-T and late-L dwarfs, and such variations can be looked for.  However, 
no one has yet provided a plausible physical explanation for this break-up 
behavior, and the inexorable thickening of the cloud 
material with decreasing \teff would seem (superficially?) to mitigate against this.
Curiously, the interplay between rapid rotation and silicate clouds for some 
critical parameter regimes might result in banding, such as 
is seen in the atmosphere of Jupiter, and this may be playing a role in the observed $J$-band brightening.
However, how such a cloud-rotation coupling could affect a transition from 
profound cloud opacity to full clarity is not immediately obvious. 
Be that as it may, the problem of fitting the latest L dwarfs 
on the infrared HR diagrams, while simultaneously also fitting their spectra,
is not explained at all by the filling factor hypothesis\footnote{Note that
we can fit the $J$ vs. $J-K$ points of the late Ls with small modal particle sizes,
but can not simultaneously reproduce their near-infrared {\it spectra} from 1.0
to 3.0 microns.  Small particle sizes ($<20$ \mic) severely flatten late L dwarf spectra in ways
not consistent with the observations.}.  
In fact, it might exacerbate it.

\begin{figure}
\epsscale{.80}
\plotone{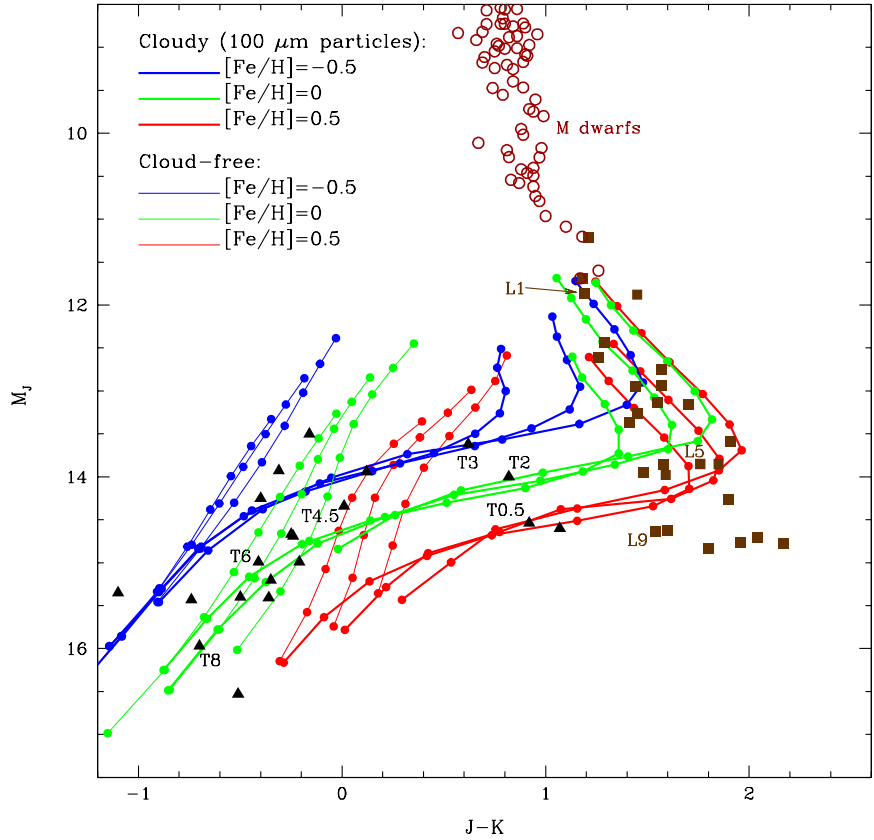}
\caption{Metallicity and gravity dependence of brown dwarf models in absolute
$J$ magnitude versus $J-K$ color space for model E parametrizations.  Cloudy models (assuming
a forsterite modal particle size of 100 \mic) and cloud-free models
(thin curves) are shown for surface gravities of
10$^{4.5}$, 10$^{5.0}$, and 10$^{5.5}$ cm s$^{-2}$ and for
metallicities, [Fe/H] = -0.5, 0, and 0.5 (blue, green, and
red curves, respectively).  Each color-coded set of 6 models
(3 cloudy and 3 cloud-free) contains the 3 different surface gravities,
with the leftmost cloudy and cloud-free curves being the highest gravity.
The points along each model curve are in \teff intervals of 100 K,
ranging from 700 K to 1500 K for the solar metallicity cloud-free models,
800 K to 1500 K for the nonsolar metallicity cloud-free models, 
900 K to 2000 K for the $g$=10$^{4.5}$ cm s$^{-2}$ solar-metallicity cloudy models,
800 K to 1900 K for the $g$=10$^{5.0}$ cm s$^{-2}$ supersolar-metallicity cloudy models,
and 800 K to 2000 K for all other cloudy models.  A difference of a
factor of 10 in metallicity carries roughly a full magnitude
difference in $J-K$ color.  Shown for comparison are L and T dwarf
data from Knapp et al. (2004) and M dwarf data from Leggett (1992).}
\label{fig:17}
\end{figure}

\begin{figure}
\epsscale{1.0}
\plotone{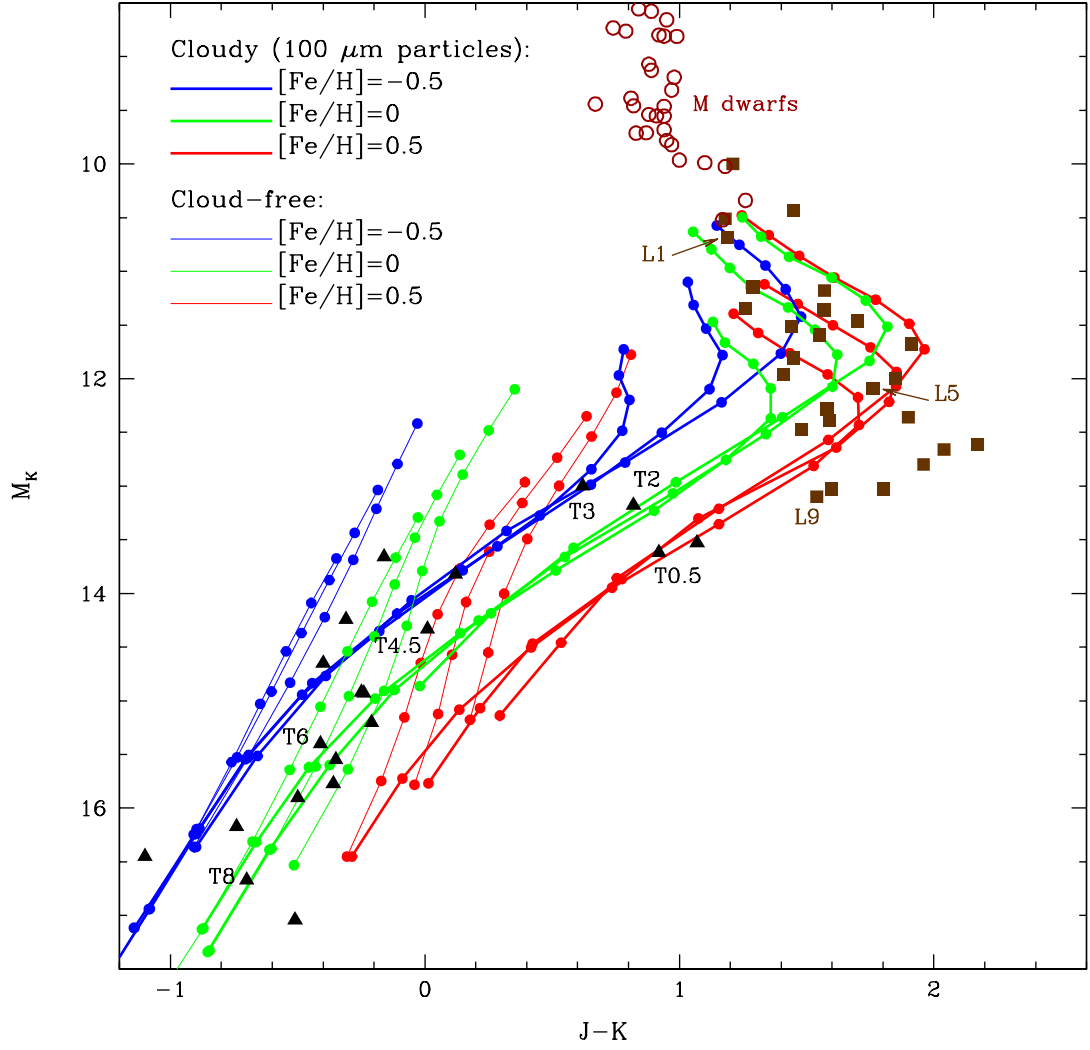}
\caption{Same as Fig. \ref{fig:17}, except in absolute
$K$ magnitude versus $J-K$ color space.}
\label{fig:18}
\end{figure}

It could be that the anomalous brightenings in $J$ and $Y/Z$ are the result of crypto-binarity.
The binary fraction of T dwarfs is not negligible and $\sim$0.75 
mag. (2.5$\times\log_{10}2$) is near the magnitude of the few 
excesses measured.  It could also be, as we show in Figs. \ref{fig:17},
that low metallicity is the culprit.  However, both explanations seem rather ad hoc
and do not explain the gaps and empty regions in the M$_J$/($J-K$) HR diagram 
currently inferred.  The redness and low $J$-band fluxes seen for the latest L dwarfs
might be explained by errors in the parallaxes or photometry. These objects might
also be explained by supersolar metallicities. However, both these 
explanations seem unlikely, particularly given the largish
number of L dwarfs that now inhabit that region of phase space.

We have shown that when calcium-aluminate, silicate, and Fe clouds first form
they do so in the radiative region, that as \teff\ decreases an isolated convective
zone emerges, and that for even lower \teffs\ the two convective zones join.  It 
remains to be seen what happens to the particle sizes and cloud morphology 
when these regions merge. Shaw (2003) and Kostinski \& Shaw  (2005) investigate the
dependence of runaway droplet growth and rainout on the presence in convective 
clouds of large velocity shears and on intermittency in the turbulence.  Could 
the merger of the outer convective cloud with the inner convective zone 
lead to regions of such large shears, in which particle growth on the 
timescales available is more rapid, and, hence, to very large particles?  
Could the merger lead to the irreversible partial flushing of cloud material into the interior? 
After the joining of the convective zones, is the timescale for grain growth too
long for the convecting feedstock to avoid being dragged into the hot interior 
before forming opaque grains?  Is a critical \teff/gravity threshold for rapid 
grain coalescence and growth reached, beyond which the average particle is too
large to contribute significant opacity (Liu, Daum, \& McGraw 2005)? Or does the scale
height of the silicate cloud collapse at some \teff\ threshold? The answers to
these questions require a multi-dimensional approach both to grain kinetics and growth
and to convective cloud structures and motions, all properly coupled.  

Since the discovery of Gleise 229B ten years ago, theorists have made great strides
in modeling brown dwarf atmospheres and spectra. The thermochemical and spectroscopic data
for the relevant molecules have improved considerably, as have the codes needed
to handle stratified atmospheres.   Observers have identified whole new spectroscopic classes,
and the observations have gotten spectacularly better across the electromagnetic spectrum.
Despite this palpable progress, much remains to be done.  Foremost are the creation 
of physically-motivated cloud models, the expansion of molecular line lists that even now remain incomplete,
the discovery of what lies beyond the T dwarfs at still lower \teffs, the further
calibration of evolutionary and spectral models with well-calibrated spectra, and the discovery
and characterization of more eclipsing systems.  Robust gravity and metallicity 
diagnostics need to be determined and pursued and the binary statistics of 
substellar-mass objects need to be put on a firm foundation. We look forward
to participating in the next generation of progress on these problematic, yet fascinating,
objects that have only recently emerged from nowhere to reawaken stellar astronomy.

\acknowledgments
 
The authors thank Mike Cushing, Adam Burgasser, Bill Hubbard, Jonathan Lunine,
Christopher Sharp, Jim Liebert, Drew Milsom, Curtis Cooper, and Jonathan Fortney
for fruitful conversations and/or help during the course of this work, as well as
NASA for its financial support via grants NAG5-10760, NNG04GL22G, and NNG05GG05G.

\clearpage

\clearpage

\clearpage

\clearpage

\clearpage

\clearpage

\clearpage

\clearpage

\clearpage

\clearpage

\clearpage

\clearpage

\clearpage

\clearpage

\clearpage

\clearpage

\clearpage

\clearpage

\clearpage

\clearpage

\begin{references}

\reference{} Ackerman, A. \& Marley, M.S. 2001, \apj, 556, 872
\reference{} Allende-Prieto, C., Lambert, D.L., \& Asplund, M. 2002, \apj, 573, L137
\reference{} Allard, F., Hauschildt, P.H., Alexander, D.R., Tamanai, A., \& Schweitzer, A. 2001, \apj, 556, 357
\reference{} Anders, E. \& Grevesse, N. 1989, Geochim. Cosmochim. Acta, 53, 197
\reference{} Borysow, A. \& Frommhold, L. 1990, \apj, 348, L41
\reference{} Borysow, A., J\o{rgensen}, U.G., \& Zheng, C. 1997, \aap, 324, 185
\reference{} Burgasser, A.J., et al.\ 1999, \apj, 522, L65
\reference{} Burgasser, A.J., et al. 2000a, \apj, 531, L57
\reference{} Burgasser, A.J., Kirkpatrick, J.\ D.,
Reid, I.\ N., Liebert, J., Gizis, J.\ E., \& Brown, M.\ E. 2000b, \aj, 120, 473
\reference{} Burgasser, A.J., et al. 2000c, \aj, 120, 1100
\reference{} Burgasser, A.J., Marley, M.S., Ackerman, A.S., Saumon, D., 
Lodders, K., Dahn, C.C., Harris, H.C., Kirkpatrick, J.D. 2004, \apj, 571, L151 
%
\reference{} Burgasser, A.J., Burrows, A., \& Kirkpatrick, J.D. 2005, submitted to \apj
\reference{} Burgasser, A.J., Reid, I.N., Kirkpatrick, J.D., Leggett, S.K., 
Liebert, J., \& Burrows, A. 2005, submitted to \apj
\reference{} Burrows, A., Marley M., Hubbard, W.B. Lunine, J.I., Guillot, T., Saumon, D.
Freedman, R., Sudarsky, D., \& Sharp, C.M. 1997, \apj, 491, 856
\reference{} Burrows, A. \& Sharp, C.M. 1999, \apj, 512, 843
\reference{} Burrows, A., Marley, M.\ S., \& Sharp, C.\ M. 2000, \apj, 531, 438
\reference{} Burrows, A., Hubbard, W.B., Lunine, J.I., \& Liebert, J. 2001, Rev. Mod. Phys., 73, 719
\reference{} Burrows, A., Ram, R.S., Bernath, P.F., Sharp,
C.M., \&\ Milsom, J.A. 2002a \apj, 577, 986
\reference{} Burrows, A., Burgasser, A.J., Kirkpatrick, J.\ D., Liebert, J., Milsom, J.A., Sudarsky, D.,
\& Hubeny, I. 2002b, \apj, 573, 394
\reference{} Burrows, A. \& Volobuyev, M. 2003, \apj, 583, 985
\reference{} Burrows, A., Sudarsky, D., \& Lunine, J. 2003, \apj, 596, 587
\reference{} Chabrier, G., Baraffe, I., Allard, F., \& Hauschildt, P. 2000, \apj, 542, 464
\reference{} Cooper, C.S., Sudarsky, D.,
Milsom, J.A., Lunine, J.I., \& Burrows, A. 2003, \apj, 586, 1320
\reference{} Cruz, K.L., Reid, I.N.,
Liebert, J., Kirkpatrick, J.D., \& Lowrance, P.J.\ 2003, \aj, 126, 2421
\reference{} Cushing, M.C., Rayner, J.T., Vacca, W.D. 2005, \apj, 623, 1115 
\reference{} Cushing, M.C., Rayner, J.T., Davis, S.P., Vacca, W.D. 2003, \apj, 582, 1066
\reference{} Dahn, C.C. et al. 2002, \aj, 124, 1170
\reference{} Deirmendjian, D. 1964, Applied Optics, 3, 187
\reference{} Delfosse, X., Tinney, C.G., Forveille, T., Epchtein, N., Bertin, E., Borsenberger, J.,
        Copet, E., De Batz, B., Fouqu\'{e}, P., Kimeswenger, S., Le Bertre, T., Lacombe, F., Rouan, D., \&
        Tiph\`{e}ne, D. 1997, \aa, 327, L25 
\reference{} Dulick, M. Bauschlicher, C.W. Jr, Burrows, A., Sharp, C.M.,
Ram, R.S. \&\ Bernath, P.F. 2003, \apj, 594, 651
\reference{} Fegley, B. \& K. Lodders, 1996, \apj, 472, L37
\reference{} Geballe, T.R., Saumon, D., Leggett, S.K., Knapp, G.R., Marley, M.S., \& Lodders, K. 2001,
\apj, 556, 373
\reference{} Geballe, T.R. et al. 2002, \apj, 564, 466
\reference{} Gelino, C.R. \& Kulkarni, S.R. 2005, astro-ph/0508164 
\reference{} Golimowski, D.A. et al. 2004, \aj, 127, 3516 
\reference{} Helling, Ch., Oevermann, M., L\"uttke, M.J.H., Klein, R., \& Sedimayr, E. 2001, \aa, 376, 194 
\reference{} Helling, Ch., Klein, R., Woitke, P, Nowak, U., \& Sedimayr, E. 2004, \aa, 423, 657 
\reference{} Hubeny, I. 1988, Computer Physics Comm., 52, 103
\reference{} Hubeny, I. 1992, in {\it The Atmospheres
of Early-Type Stars}, ed. U. Heber \& C. J. Jeffery, Lecture
Notes in Phys. 401, (Berlin: Springer), p. 377
\reference{} Hubeny, I. \& Lanz, T. 1995, \apj, 439, 875
\reference{} Hubeny, I., Burrows, A., \& Sudarsky, D. 2003, \apj, 594, 1011
\reference{} Jones, H.R.A. \& Tsuji, T. 1997, \apj, 480, L39
\reference{} Kirkpatrick, J.D., Beichman, C.A., \& Strutskie, M.F.
1997, \apj, 476, 311
\reference{} Kirkpatrick, J.D., Reid, I.N., Liebert, J., Cutri, R.M., Nelson, B.,
                   Beichman, C.A., Dahn, C.C., Monet, D.G., Gizis, J., \& Skrutskie, M.F. 1999,
                   \apj, 519, 802
\reference{} Kirkpatrick, J.\ D., Reid, I.\ N.,
Liebert, J., Gizis, J.\ E., Burgasser, A.\ J., Monet, D.\ G., Dahn, C.\ C.,
Nelson, B., \& Williams, R.\ J. 2000, \aj, 120, 447
\reference{} Knapp, G. et al. 2004, \aj, 127, 3553 
%
\reference{} Kostinski, A.B. \& Shaw, R.A. 2005, Bull. Amer. Met. Soc., February issue 2005, p. 235
\reference{} Leggett, S.K. 1992, \apj~Suppl., 82, 351
\reference{} Leggett, S.K. et al. 2000, \apj, 536, L35
\reference{} Leggett, S.K., Allard, F., Geballe, T.R.,
Hauschildt, P.H, \& Schweitzer, A. 2001 \apj, 548, 908
\reference{} Liu, M.C. \& Leggett, S.K. 2005, \apj, in press (astro-ph/0508082) 
\reference{} Liu, Y., Daum, P.H., \& McGraw, R.L. 2005, Geophys. Res. Letters, 32, L11811
\reference{} Lodders, K. 1999, \apj, 519, 793
\reference{} Marley, M.S., Saumon, D., Guillot, T., Freedman, R.S., Hubbard, W.B.,
Burrows, A. \& Lunine, J.I. 1996, Science, 272, 1919
\reference{} Marley, M.S., Seager, S., Saumon, D., Lodders, K., Ackerman, A.S., Freedman, R., \& Fan, X.
2002, \apj, 568, 335 
\reference{} Marley, M.S., Cushing, M.C., \& Saumon, D. 2004, in the proceedings of the 13th 
Cambridge Workshop on Cool Stars, Stellar Systems, and the Sun, Hamburg, Germany, 
July 2004, astro-ph/0409267
\reference{} Mart{\'{\i}}n, E.\ L., Delfosse, X.,
Basri, G., Goldman, B., Forveille, T., \& Zapatero Osorio, M.\ R. 1999, \aj, 118, 2466
\reference{} McLean, I.S., McGovern, M.R., Burgasser, A.J., Kirkpatrick, J.D., Prato, L., Kim, S.S. 2003, \apj, 596, 561
\reference{} Mihalas, D. 1978, {\it Stellar
Atmospheres, 2nd Ed.}, (San Francisco: Freeman \& Co.)
\reference{} Nakajima, T., Tsuji, T., \& Yanagisawa, K. 2004, \apj, 607, 499
\reference{} Oppenheimer, B.R., Kulkarni, S.R., Matthews, K., \& Nakajima, T. 1995,
                  Science, 270, 1478
\reference{} Patten, B.M. et al. 2004,  BAAS, 205, 1110
\reference{} Partridge, H. \& Schwenke, D.W. 1997,
J. Chem. Phys., 106, 4618
\reference{} Saumon, D., Chabrier, G.,
\& Van Horn, H.M. 1995, \apjs, 99, 713
\reference{} Shaw, R.A. 2003, Annu. Rev. Fluid Mech., 35, 183
\reference{} Sudarsky, D., Burrows, A., \& Hubeny, I. 2003, \apj, 588, 1121
\reference{} Tinney, C.G., Burgasser, A.J., \& Kirkpatrick, J.D. 2003, \aj, 126, 975 
%
\reference{} Tokunaga, A.T. \& Simons, D.A. 2003, Instrument Design and 
Performance for Optical/Infrared Ground-based Telescopes, edited by Iye, 
Masanori and Moorwood, Alan F.M., in the Proceedings of the SPIE, Vol. 4841, pp. 420-424 
\reference{} Tokunaga, A.T., Simons, D.A., \& Vacca, W.D. 2002, PASP, 114, 180
\reference{} Tsuji, T., Ohnaka, K., \& Aoki, W. 1999, \apj, 520, L119
\reference{} Tsuji, T., \& Nakajima, T. 2003, \apj, 585, L151 
\reference{} Tsuji, T., Nakajima, T., \& Yanagisawa, K. 2004, \apj, 607, 511 
\reference{} Tsvetanov, Z.I. et al. 2000, \apj, 531, 61
\reference{} Vrba, F. J. et al. 2004, \aj, 127, 2048 
\reference{} Woitke, P. \& Helling, Ch. 2003, \aa, 399, 297 
\reference{} Woitke, P. \& Helling, Ch. 2004, \aa, 414, 335 

\end{references}
\end{document}